\PassOptionsToPackage{dvipsnames}{xcolor}
\PassOptionsToPackage{nounderscore}{syntax}

\documentclass{article}
\usepackage[preprint]{neurips_2025}

\usepackage[utf8]{inputenc} 
\usepackage[T1]{fontenc}    
\usepackage{amsfonts}       
\usepackage{amssymb}
\usepackage{amsthm}
\usepackage{nicefrac}       
\usepackage{microtype}      

\usepackage{tikz}
\usepackage{amsmath}
\usepackage{graphicx}
\usepackage{url}
\usepackage{color}
\usepackage{multicol}
\usepackage{amsmath}
\usepackage{tikz}
\usetikzlibrary{shapes.geometric, shapes.arrows, shapes.misc, quotes, positioning, shapes, chains, calc, arrows.meta}
\usepackage{xspace}
\usepackage{enumitem}
\usepackage[ruled,linesnumbered,noend,vlined]{algorithm2e}
\usepackage{rotating}
\usepackage{multirow}
\usepackage{booktabs}

\usepackage{makecell}

\usepackage{stmaryrd}

\usepackage{bm}

\usepackage{adjustbox}

\usepackage{subfig}

\usepackage{mathtools}

\usepackage{pgfplots}
\pgfplotsset{compat=1.18}

\newcommand{\eat}[1]{}
\newcommand{\eg}{e.g.,\xspace}
\newcommand{\ie}{i.e.,\xspace}

\newcommand{\first}{\textsf{(i)}\xspace}
\newcommand{\second}{\textsf{(ii)}\xspace}
\newcommand{\third}{\textsf{(iii)}\xspace}

\newcommand{\sys}{\textsf{BlockA2A}\xspace}

\usepackage{listings}

\usepackage{float}
\usepackage{newfloat}

\usepackage{setspace}
\usepackage{mathtools,balance}

\newtheorem{theorem}{Theorem}

\usepackage[capitalize,nameinlink]{cleveref}
\crefname{BNF}{Grammar}{Grammars}

\newcommand{\paraspace}{\vspace{0.01in}}
\newcommand{\parab}[1]{\paraspace\noindent{\bf #1.}}

\usepackage{newfloat}
\usepackage{caption}

\DeclareFloatingEnvironment[fileext=frm,placement={!htbp},name=Functionality]{Functionality}
\DeclareFloatingEnvironment[fileext=frm,placement={!htbp},name=Protocol]{Protocol}
\DeclareFloatingEnvironment[fileext=frm,placement={!htbp},name=Application]{Application}
\usepackage{placeins}
\PassOptionsToClass{table, xcdraw}{xcolor}
\definecolor{lbcolor}{rgb}{0.95,0.95,0.95}

\lstdefinelanguage{json}{
    basicstyle=\ttfamily,
    keywordstyle=\color{blue}\bfseries,
    stringstyle=\color{brown},
    commentstyle=\color{gray},
    showstringspaces=false,
    morestring=[b]",
    morecomment=[l]{//},
    morekeywords={id,publicKey,type,publicKeyMultibase,service,serviceEndpoint,capabilities,supportedModels,maxComputeTime,permissions,policy-constraints,allowed_interaction_hours,max_data_size,proof,created,verificationMethod,proofValue}
}

\usepackage[framemethod=TikZ]{mdframed}
\usepackage{lipsum}

\newdimen\linenumbersep

\newcommand{\linenumber}[1]{%
  \linenumbersep 4pt%
  \advance\linenumbersep\mdflength{innerleftmargin}%
  \advance\linenumbersep\mdflength{innerlinewidth}%
  \advance\linenumbersep\mdflength{middlelinewidth}%
  \advance\linenumbersep\mdflength{outerlinewidth}%
  \advance\linenumbersep\mdflength{linewidth}%
  \makebox[0pt][r]{{\rmfamily\tiny#1}\hspace*{\linenumbersep}}}

\usepackage{dashrule}

\mdfdefinestyle{FunctionalityFrame}{%
    linecolor=black,
    outerlinewidth=0.5pt,
    font=\normalsize,
    roundcorner=2pt,
    innertopmargin=2pt,
    innerbottommargin=2pt,
    innerrightmargin=4pt,
    innerleftmargin=3pt,
    backgroundcolor=lbcolor,
    nobreak=true
    }

\mdfdefinestyle{ProtocolFrame}{%
    linecolor=black,
    outerlinewidth=0.5pt,
    font=\normalsize,
    roundcorner=1pt,
    innertopmargin=2pt,
    innerbottommargin=2pt,
    innerrightmargin=4pt,
    innerleftmargin=3pt,
    nobreak=true
    }

\mdfdefinestyle{ApplicationFrame}{%
    linecolor=black,
    outerlinewidth=0.5pt,
    font=\normalsize,
    roundcorner=1pt,
    innertopmargin=2pt,
    innerbottommargin=2pt,
    innerrightmargin=4pt,
    innerleftmargin=3pt,
    nobreak=true
    }

\usepackage{algpseudocode}

\usepackage{tabularx}
\usepackage{array,etoolbox}
\preto\tabular{\setcounter{magicrownumbers}{0}}
\newcounter{magicrownumbers}

\usepackage{booktabs}
\usepackage{pifont}
\usepackage[flushleft]{threeparttable}

\newcounter{protocol}


\usepackage[sets, operators]{cryptocode}

\author{%
  Zhenhua Zou \\
  Tsinghua University\\
  \texttt{zou-zh21@mails.tsinghua.edu.cn} \\
  \And
  Zhuotao Liu \thanks{Corresponding author.} \\
  Tsinghua University \\
  \texttt{zhuotaoliu@tsinghua.edu.cn} \\
  \AND
  Lepeng Zhao \\
  Tsinghua University \\
  \texttt{zhaolp22@mails.tsinghua.edu.cn} \\
  \And
  Qiuyang Zhan \\
  Tsinghua University \\
  \texttt{zhanqy24@mails.tsinghua.edu.cn} \\
}

\begin{document}

\title{\sys: Towards Secure and Verifiable Agent-to-Agent Interoperability \thanks{Position Paper: under active development.}}

\maketitle

\begin{abstract}
    The rapid adoption of agentic AI, powered by large language models (LLMs), is transforming enterprise ecosystems with autonomous agents that execute complex workflows. Yet we observe several key security vulnerabilities in LLM-driven multi-agent systems (MASes): fragmented identity frameworks, insecure communication channels, and inadequate defenses against Byzantine agents or adversarial prompts. In this paper, we present the first systematic analysis of these emerging multi-agent risks and explain why the legacy security strategies cannot effectively address these risks. Afterwards, we propose \sys, the first unified multi-agent trust framework that enables secure and verifiable and agent-to-agent interoperability. At a high level, \sys adopts decentralized identifiers (DIDs) to enable fine-grained cross-domain agent authentication, blockchain-anchored ledgers to enable immutable auditability, and smart contracts to dynamically enforce context-aware access control policies. \sys eliminates centralized trust bottlenecks, ensures message authenticity and execution integrity, and guarantees accountability across agent interactions. Furthermore, we propose a Defense Orchestration Engine (DOE) that actively neutralizes attacks through real-time mechanisms, including Byzantine agent flagging, reactive execution halting, and instant permission revocation.
    
    Empirical evaluations demonstrate \sys's effectiveness in neutralizing prompt-based, communication-based, behavioral and systemic MAS attacks. We formalize its integration into existing MAS and showcase a practical implementation for Google's A2A protocol. Experiments confirm that \sys and DOE operate with sub-second overhead, enabling scalable deployment in production LLM-based MAS environments.
\end{abstract}

\pagestyle{plain}

\section{Introduction}
\label{sec:intro}

The emergence of agentic AI, powered by large language models (LLMs), represents a transformative leap in AI capabilities, enabling autonomous agents to execute complex tasks, adapt to dynamic real-world environments, and collaborate seamlessly with humans and other agents \cite{agentic-ai-survey, role-of-agentic-ai, react, webarena, mind2web}. These agents are reshaping industries by automating workflows, enhancing decision-making, and unlocking efficiencies of cross-system collaboration. Gartner predicts that by 2029, agentic AI will independently handle 80\% of common customer service issues, cutting human intervention by 70\%—demonstrating its growing decision-making autonomy \cite{gartner}. Recent IDC reports highlights scaled deployment of agentic AI in key business functions of top enterprises \cite{idc-agenticai}. Forbes adds that agentic AI is now core B2B infrastructure, automating supply chain collaboration, contract negotiations, and beyond~\cite{forbes-agenticai}.

This shift toward interconnected autonomy hinges on \textit{agent-to-agent collaboration}~\cite{ioa, a2a-protocol, chateval, autogen, debate, langgraph-multi-agent}, a cornerstone of the agentic era. 
By coordinating across silos (such as integrating customer service agents with real-time data analysis tools or supply chain optimizers), these agents dynamically share insights, tasks, and expertise to resolve cross-departmental challenges or optimize multi-step processes \cite{autogen, camel, metagpt}. Such synergy not only automates end-to-end workflows but also fosters adaptive problem-solving, leveraging collective intelligence to address complexities that single agents cannot manage alone. The result is a scalable and resilient multi-agent system (MAS) that amplifies operational efficiency and drives innovation, positioning agentic AI as the backbone of modern enterprise ecosystems.

However, the agent-to-agent interoperability 
also introduces unprecedented security vulnerabilities \cite{autodefense, agentguard}. First, the absence of a universal trust framework leaves ecosystem fragmented: agents from different developers/organizations often operate on mismatched security standards, making it difficult to validate identities, verify message authenticity and trace data authorship across collaborative tasks. 
Second, the intricate network of inter-agent communications—spanning APIs, real-time data exchanges, and shared interfaces—creates a sprawling attack surface, vulnerable to exploits like data interception, command injection, or workflow sabotage (e.g., manipulating supply chain agents to falsify inventory updates) \cite{jailbreak-survey, achilles-heel, breaking-agents, multi-agent-exe-code, collaboration-attack}. 
Third, malicious ``Byzantine'' agents, whether compromised or intentionally adversarial, can disrupt workflows, poison shared data, or exfiltrate sensitive information \cite{blockagents}, while LLM-driven collaboration introduces unique risks: adversarial prompts could hijack decision-making processes, and low-quality or toxic outputs from one agent may propagate unchecked across networks, corrupting downstream actions \cite{agent-smith, agents-under-siege, prompt-infection, security-tax}. These challenges demand urgent innovation in universal trust frameworks, dynamic authentication/access-control protocols, and AI-native safeguards to ensure trust, accountability, and resilience in an era where autonomous systems hold the keys to enterprise success.

Despite these escalating risks, current agent-to-agent frameworks (including protocols like Google's A2A~\cite{a2a-protocol,a2a-analysis}) fail to align security strategies with the critical challenges outlined~\cite{autogen, camel, crewai, langgraph, metagpt}. 
In particular, centralized identity authentication models, as seen in A2A, create single points of failure and struggle to verify cross-domain agent identities securely. Data integrity mechanisms relying on HTTPS or OAuth lack safeguards for long-term tamper-proof verification, leaving historical interactions vulnerable to manipulation. Audit trails are equally brittle: centralized logging systems are prone to tampering or gaps, while inconsistent logs across different organizations (in the same multi-agent workflow) hinder accountability. Further, static permission controls in dynamic environments often lag behind real-time needs, resulting in over-provisioned access or delayed revocation—a flaw exploited by malicious actors to escalate privileges. These mismatches not only expose agentic ecosystems to Byzantine behaviors and prompt-based attacks but also amplify systemic vulnerabilities, as compromised trust or corrupted data propagates unchecked (see our detailed discussion in \S~\ref{sec:literature-review}). Addressing these gaps demands novel frameworks that prioritize decentralized trust, immutable auditability, and granular, context-aware access controls to secure the future of agentic collaboration.

Toward this end, we present \sys, a unified trust framework designed to safeguard all paradigms of agent-to-agent collaboration while flexibly enabling diverse defense strategies. At its core, \sys addresses the shortcomings of traditional security frameworks—centralized trust base, audit challenges, and coarse-grained permissions—by integrating three core architectural pillars: decentralized identity, immutable ledger, and smart contract enforcement. As \sys's foundation, the \textit{Identity Layer} leverages decentralized identifiers (DID) and cryptographic authentication to eliminate single points of trust, enabling seamless cross-domain agent verification without centralized authorities (\S~\ref{subsec:identity-layer}). Complementing this, the \textit{Ledger Layer} ensures tamper-proof auditability: critical interaction data (e.g., task player identities, task inputs/outputs, state transitions) is anchored to blockchain via Merkle proofs, reconciling multi-agent log inconsistencies and guaranteeing non-repudiation (\S~\ref{subsec:ledger-layer}). Finally, the \textit{Smart Contract Layer} embeds dynamic, context-aware policies through smart contracts—automating granular access control (e.g., revoking compromised agents in real-time) and enforcing collaboration logic (e.g., validating prompt integrity before execution) (\S~\ref{subsec:smart-contract-layer}). By unifying the three layers, \sys not only closes drawbacks in legacy agent-to-agent security frameworks, but also provides a scalable, extensible foundation for implementing defense-in-depth strategies—from Byzantine fault tolerance to adversarial prompt detection—ensuring agentic AI systems remain secure, accountable, and resilient in the face of growing system complexity and evolving threats.

We further propose adaptive trust integration and threat-aware modularity to extend \sys's capabilities. First, \sys can be instantiated across diverse Multi-Agent Systems (MASes)—including Supervisor-Based, Network/Graph-Based, and Federated Learning-Based models—through a formal framework (\S~\ref{sec:instantiation-framework}). This framework employs rigorous protocol translation to map MAS-specific data and protocols to \sys's canonical formats, enables modular and pluggable integration for selective layer adoption, and ensures trust preservation across MASes. Second, a Defense Orchestration Engine (DOE) leverages \sys's three-layer architecture for proactive threat detection, automated response, and forensic analysis (\S~\ref{sec:doe}). The deep integration with \sys's Identity, Ledger, and Smart Contract layers allows the DOE to use DIDs for authentication, monitor on-chain events for integrity, and dynamically update smart contracts for real-time policy enforcement. By unifying these designs, \sys evolves into an active trust substrate, fortifying ecosystems against threats while maintaining the flexibility crucial for open, cross-domain collaboration.

We thoroughly evaluates \sys's effectiveness and efficiency in safeguarding MASes against diverse threats (\S~\ref{sec:evaluation}). Our empirical analysis demonstrates \sys's robust defense capabilities against prompt-based, communication-based, behavioral/psychological, and systemic/architectural attacks, leveraging its unique layered architecture and the Defense Orchestration Engine (DOE). Furthermore, we provide a detailed instantiation of \sys within Google A2A~\cite{a2a-protocol}, showcasing its practicality and ability to enhance authenticity, integrity, and accountability without disrupting existing protocols. Crucially, our operational cost analysis reveals that \sys introduces reasonable overheads, with most critical security operations completing within sub-second timeframes, proving its viability for real-time defense in complex MAS environments.

In summary, our work advances agent collaboration security and trustworthiness through three key contributions:

\begin{itemize}

\item First Systematic Agent-to-Agent Security Analysis: We conduct the first comprehensive review of agent-to-agent collaboration paradigms, threats, and defenses, mapping the security landscape through rigorous taxonomy. By exposing critical mismatches between evolving threats (e.g., Byzantine agents, adversarial prompts) and outdated security strategies (e.g., centralized trust, fragmented defenses and poorly aligned logs), we demonstrate the urgent need for a unified trust framework to safeguard agent-to-agent interoperability.
\item \sys Framework: To address these problems, we introduce \sys, the first unified framework combining decentralized identity, immutable ledger, and smart contract enforcement to resolve legacy vulnerabilities: single points of failure, auditability, and coarse-grained/delayed access controls. Enhanced by adaptive trust integration and threat-aware modularity, \sys offers a flexible, extensible foundation for securing MASes with diverse collaboration paradigms—from agent networks to federated learning—while enabling defense-in-depth strategies
\item Empirical Validation of Effectiveness and Efficiency: Our empirical studies confirm its robust defense against diverse attack vectors, from prompt injection to systemic exploits. We demonstrate \sys's practical applicability through a detailed instantiation within Google A2A, showcasing its seamless integration for enhanced authenticity, integrity, and accountability. Critically, our evaluation highlights \sys's real-time responsiveness, with crucial security operations completing within sub-second intervals, proving its viability for proactive defense in dynamic MASes. 
\item Open Source Release: We have open sourced our entire implementation of \sys. The codebase, documentation, and integration examples are publicly available on GitHub\footnote{\url{https://github.com/Jacobzqy/BlockA2A}}, enabling practitioners and researchers to reproduce our results, extend our defenses, and accelerate the adoption of secure and verifiable agent-to-agent collaboration.
\end{itemize}
\section{Multi-Agent System: Definitions, Frameworks, Attacks and Defenses}
\label{sec:literature-review}

This section starts by clearly defining multi-agent systems (MASes), including agents, their states, actions, communication languages, and interaction protocols. It explains how autonomous agents coordinate through structured message exchanges controlled by finite-state machine protocols, enabling synchronized state changes and collaboration. We then review key MAS frameworks and collaboration patterns, outline common security threats, and assess current defenses and their limitations. Our analysis highlights the urgent need for unified identity management, stronger trust in interactions, and improved access control in complex multi-agent environments.

\subsection{Definitions}
\label{subsec:definitions}

A multi-agent system (MAS) constitutes the systematic exchange of structured information between autonomous computational entities, enabling coordination, collaboration, or resource sharing. Formally, let $ \mathcal{A} = \{a_1, a_2, \dots, a_n\} $ represent a finite set of agents, where each agent $ a_i \in \mathcal{A} $ is characterized by:  
\begin{itemize}
    \item A state space $ \mathcal{S}_i $, with $ s_i(t) \in \mathcal{S}_i $ denoting its state at time $ t $,  
    \item An action space $ \mathcal{A}_i $ defining permissible interactions with the environment (including message generation),  
    \item A communication language $ \mathcal{L} = \langle \Sigma, \mathcal{M} \rangle $, consisting of a syntax $ \Sigma $ (symbol set) and semantics $ \mathcal{M} $ (interpretation rules).  
\end{itemize}

A communication message $ m \in \mathcal{M} $ is a ternary tuple:  
$$
m = \langle \text{sender}(m), \text{receiver}(m), \text{content}(m) \rangle
$$  
where $ \text{sender}(m), \text{receiver}(m) \in \mathcal{A} $ denote the message originator and target, and $ \text{content}(m) \in \mathcal{L} $ specifies the encoded information.  

Communication protocols governing message exchange are modeled as finite-state machines (FSMs):  
$$
\mathcal{P} = \langle \mathcal{Q}, q_0, \mathcal{T}, \mathcal{F} \rangle
$$  
with:  
\begin{itemize}
    \item $ \mathcal{Q} $: finite set of protocol states,  
    \item $ q_0 \in \mathcal{Q} $: initial state,  
    \item $ \mathcal{T} \subseteq \mathcal{Q} \times \mathcal{M} \times \mathcal{Q} $: transition relation (dictating allowable messages at each protocol state),  
    \item $ \mathcal{F} \subseteq \mathcal{Q} $: set of final (accepting) states.  
\end{itemize}

The transition relation $ \mathcal{T} $ constrains message exchange to sequences compliant with the protocol. Agents generate messages based on their internal states $ s_i(t) $ and action spaces $ \mathcal{A}_i $, which must align with the protocol's current state $ q \in \mathcal{Q} $.  

An interaction trace $ T = \{m_1, m_2, \dots, m_k\} $ conforms to protocol $ \mathcal{P} $ iff there exists a state sequence $ q_0 \xrightarrow{m_1} q_1 \xrightarrow{m_2} \dots \xrightarrow{m_k} q_k $ such that $ (q_{i-1}, m_i, q_i) \in \mathcal{T} $ for all $ i \in [1,k] $. Each message $ m_i $ simultaneously advances the protocol state $ q_{i-1} \to q_i $ and triggers a state transition in the receiving agent $ a_j $, establishing synchronization between protocol progression and agent behavior.  

Each agent processes messages via a state transition function $ \delta_i: \mathcal{S}_i \times \mathcal{M} \rightarrow \mathcal{S}_i $, updating its state as:  
$$
s_i(t+1) = \delta_i(s_i(t), m(t))
$$  
upon receiving message $ m(t) $. This update occurs in tandem with the protocol's transition to a new state, creating a bidirectional coupling: protocol states govern permissible messages, which in turn drive agent state transitions, while agent states determine subsequent messages compliant with the protocol.  

This formal framework establishes the foundation for multi-agent systems (MASes), where distributed decision-making relies on structured information flow governed by syntactic and semantic rules, with protocol and agent states co-evolving through message exchange.

\subsection{Existing MAS Frameworks and Protocols}\label{sec:frameworks}

The development of sophisticated LLM-based MASes relies heavily on the underlying frameworks and protocols that facilitate communication and coordination between individual agents. Several noteworthy frameworks and protocols have emerged, each with its own unique approach to enabling agent collaboration, ranging from comprehensive software development kits to standardized communication specifications.

\subsubsection{Patterns of Multi-Agent Collaboration}

\parab{Execution Graph/Network-Based Collaboration}
Frameworks in this category model agent interactions as directed graphs or networks, emphasizing workflow orchestration through topological structures and stateful transitions.

\textit{LangGraph} exemplifies this pattern through its explicit graph-based architecture, where agent workflows are modeled as directed graphs, enabling developers to define complex stateful interactions with fine-grained control over execution flows. The framework's support for durable execution and comprehensive memory management (both short-term and persistent) ensures reliable long-running collaborations. Its integration with LangSmith provides visualizations of execution paths and state transitions, making it ideal for debugging network-based agent interactions~\cite{langgraph}. 

The \textit{BotSharp} framework implements pipeline flow execution through its plug-in architecture and "Routing \& Planning" module. This enables sequential or parallel task execution across agents while maintaining individual states. The .NET-based framework’s modular design allows enterprises to construct execution networks where agents with specialized roles process information through predefined pipelines~\cite{botsharp}.

The \textit{OpenAI Agents SDK} facilitates dynamic execution networks via its "Handoffs" mechanism. Agents configured with specialized tools and instructions can transfer control to peers through explicit handoff operations, creating ad-hoc execution chains. The SDK’s provider-agnostic design allows these networks to leverage heterogeneous LLMs, while built-in tracing capabilities map the emergent collaboration graph~\cite{openai-agent-sdk}.

\textit{CAMEL AI} demonstrates research-focused patterns through projects like OWL (Optimized Workforce Learning) combining execution graphs with reinforcement learning, and CRAB (Cross-environment Agent Benchmark) blending network-based collaboration with multimodal evaluation. Its MCP (Model Context Protocol) integration enables mixed supervisory/network architectures connecting diverse data sources~\cite{camel}.

\parab{Supervisor-Based Collaboration (Hierarchical)}
These frameworks employ coordination mechanisms—hierarchical or role-based—to manage agent collaboration without strict graph structures, often through centralized or distributed supervision.

\textit{MetaGPT} operationalizes hierarchical supervision by simulating software company roles. A supervisory layer (e.g., product managers/architects) decomposes requirements into sub-tasks, which are then assigned to specialized role agents (engineers/testers). This hierarchy is governed by encoded Standard Operating Procedures (SOPs), ensuring structured coordination reminiscent of human organizational workflows~\cite{metagpt}.

The \textit{Agent Development Kit (ADK)} supports hierarchical control through its workflow agent types. Sequential agents enforce stepwise task progression, while Loop agents implement supervisor-driven iteration protocols. The A2A communication standard~\cite{a2a-protocol} enables supervisors to monitor subordinate agents’ states and intervene when predefined conditions are met~\cite{adk}.

\textit{smolagents} explicitly advertises support for "multi-agent hierarchies," though technical specifics remain undocumented. The framework’s barebones design suggests lightweight implementation of supervisory control patterns where parent agents coordinate child agents through simplified orchestration rules~\cite{smolagents}.

\textit{AWS Agent Squad} (formerly Multi-Agent Orchestrator) introduces \textit{SupervisorAgent}, a centralized coordinator that implements an "agent-as-tools" architecture. The SupervisorAgent dynamically routes queries, delegates subtasks to specialized agents (e.g., Bedrock/Lex), and maintains conversation context across parallel workflows. Its support for dual-language implementation (Python/TypeScript), DynamoDB storage, and CloudWatch monitoring reinforces its role as a hierarchical supervisor~\cite{agent-squad}.

\parab{Federated Learning-Based Collaboration}
Although no existing frameworks explicitly implement federated learning paradigms~\cite{yq-fl,martfl} for multi-agent collaboration, this pattern presents theoretical potential for decentralized knowledge synthesis across agents. A federated approach could enable agents with private data silos to collaboratively refine shared models while preserving data sovereignty. In particular, horizontal federated learning could enhance homogeneous agent models owned by multiple organizations through gradient aggregation, while vertical federated learning allows the synthesis of multi-modal features. Technical challenges include developing: \first Secure parameter aggregation protocols across untrusted agents; \second Privacy-preserving guarantees during model aggregation and updates; \third Consensus mechanisms for knowledge integration and reward distribution.

\subsection{Attacks}

While MASes leverage distributed intelligence to tackle complex tasks and deliver resilient services, their collaborative architecture inherently expands systemic attack surfaces through emergent vulnerabilities. Single-agent weaknesses—ranging from adversarial prompts to jailbreaking—are amplified in MAS via error propagation and cross-contamination risks across interconnected agents. This subsection establishes an attack taxonomy across four classes—prompt-based, communication-based, behavioral/psychological, and systemic/architectural attacks—synthesizing findings from cutting-edge MAS security research. 

\parab{Prompt-Based Attacks}
Prompt-based attacks manipulate the input prompts provided to LLMs within MAS to induce malicious behavior. These attacks exploit the language model's sensitivity to input phrasing to bypass safety constraints or generate harmful outputs. For example, \cite{jailbreak-survey} introduces jailbreaking attacks, which use adversarial prompts to compel LLMs to generate responses that violate usage policies or ethical guidelines. Similarly, \cite{g-safeguard} highlights adversarial prompt injection as a method to trigger misinformation propagation or unintended system behaviors. A notable variant is \textit{Prompt Infection}, where malicious prompts self-replicate across interconnected agents, analogous to a computer virus \cite{netsafe, prompt-infection}. This attack can lead to data theft, scams, and system-wide disruption while remaining stealthy.

\parab{Communication-Based Attacks}
Communication-based attacks target the interaction channels between agents, disrupting information flow or injecting false data. \cite{red-team-comm} proposes the Agent-in-the-Middle (AiTM) attack, which intercepts and modifies inter-agent messages, exploiting fundamental communication mechanisms in LLM-MAS. Another example is the false-data injection attack described in \cite{ripple-effect}, where attackers compromise communication links to destabilize leader-following control processes. Additionally, \cite{corba} introduces contagious recursive blocking attacks, which disrupt information exchange by forcing agents into repetitive or irrelevant actions, reducing system availability.

\parab{Behavioral/Psychological Attacks}
Behavioral or psychological attacks exploit the decision-making processes of agents, particularly those influenced by simulated personality traits or psychological models. \cite{psysafe} identifies how dark personality traits in agents can lead to risky behaviors and evaluates MAS safety from psychological and behavioral perspectives. These attacks aim to manipulate agents into making suboptimal decisions or engaging in harmful actions. For instance, \cite{breaking-agents} presents malfunction amplification attacks, which mislead agents into executing repetitive or irrelevant actions, causing system malfunctions.

\parab{Systemic/Architectural Attacks}
Systemic or architectural attacks target the underlying infrastructure or design of MAS, leveraging topological vulnerabilities or systemic weaknesses. \cite{g-safeguard} introduces the G-Safeguard framework, which addresses topological vulnerabilities that enable attacks like adversarial misinformation propagation. \cite{achilles-heel} identifies critical trustworthiness challenges in distributed MAS, including susceptibility to malicious attacks, communication inefficiencies, and system instability. These attacks can degrade system performance by up to 80\% and exploit vulnerabilities across various frameworks \cite{achilles-heel}. Additionally, \cite{agents-under-siege} presents optimized prompt attacks that bypass distributed safety mechanisms by exploiting latency and bandwidth constraints in network topologies.


\subsection{Defenses and Mismatches}

While the proliferation of LLM-based multi-agent systems (MAS) has spurred the development of defensive strategies in response of the aforementioned attacks, current approaches exhibit significant mismatches with the emerging threat landscape. This subsection analyzes existing defenses and highlights critical gaps in addressing MAS-specific vulnerabilities.

\subsubsection{Existing Defenses}

Current defensive mechanisms can be broadly categorized into three types:

\parab{Attack-Specific Defenses} 
These reactive measures, such as prompt filtering \cite{g-safeguard} and adversarial training \cite{trustworthy-survey}, are designed to counter known attack vectors (e.g., jailbreak prompts or malicious code injection). However, their efficacy is limited by their inability to anticipate novel threats, rendering them non-forward-looking and susceptible to zero-day exploits.

\parab{Framework-Equipped Defenses} 
Traditional network trust and security strategies, such as Certificate Authority (CA), TLS, intrusion detection systems and firewalls, are often integrated into MAS frameworks \cite{cut-the-crap}. While these mechanisms address low-level network vulnerabilities, they inadequately protect against content/data-based attacks (e.g., prompt injection, misinformation propagation) that target the semantic layer of agent interactions. Recent MAS frameworks integrate lightweight security mechanisms tailored to their architectures, in particular:
\begin{itemize}
 \item CAMEL enforces role boundaries via system prompts (e.g., prohibiting role flipping) and uses conversation termination triggers (e.g., $<TASK\_DONE>$ token) to prevent loops \cite{camel}.
 \item MetaGPT employs structured communication (e.g., PRDs, API specs) and a publish-subscribe message pool to filter interactions by role, though this lacks content integrity checks \cite{metagpt}.
 \item AutoGen introduces a Safeguard Agent for code auditing and human-in-the-loop validation, but these rely on heuristic rules \cite{autogen}.
 \item CrewAI uses role-based task hierarchies and component fingerprinting for auditability \cite{crewai}.
 \item LangGRAPH enables human-paused workflows and state persistence for fault tolerance \cite{langgraph}.
 \item ADK applies guardrails (e.g., tool context enforcement, content filters, safety callbacks), sandboxed execution (e.g., Vertex Code Interpreter), VPC-SC perimeters and audit tracing \cite{adk}.
 \item ANP provides human authorization for high‑risk operations, enforces permission isolation via hierarchical key management and dynamic verification, employs a multi‑DID strategy with rotated least‑privilege sub‑DIDs, and ensures minimal disclosure through end‑to‑end ECDHE encryption \cite{anp-did}.
\end{itemize}
However, these mechanisms universally suffer from \first over-reliance on prompt engineering (vulnerable to jailbreak attacks); \second lack of cryptographic trust layers (e.g., data and interaction tracing); \third inability to handle dynamic cross-domain access control in heterogeneous agent ecosystems.

\parab{Blockchain-Enabled Coordination} 
Approaches like BlockAgents \cite{blockagents} leverage blockchain to facilitate Byzantine-robust decision-making in MAS. However, these solutions primarily focus on decisional consensus and fail to fully exploit blockchain's potential for establishing verifiable agent identities, tracing interaction histories, or enforcing accountability for content/data origins.

\subsubsection{Unveiled Gaps}

We summarize the key gaps between exisiting defenses and the urgent MAS vulnerabilities below:

\parab{Absence of Universal Identity Mechanisms} 
The emergence of diverse agent collaboration paradigms (\eg supervisor-based, network/graph-based) and heterogeneous development ecosystems necessitates cross-domain identity verification and resource sharing. Current frameworks rely on centralized certificate authorities (CAs), which are ill-suited for decentralized and fragmented MAS systems. While solutions like ANP \cite{anp-did} propose decentralized identifiers (DIDs) for agent identity management, their cross-domain interoperability remains underdeveloped. Besides, it is unclear how to seamlessly migrate identities in legacy systems to DID, limiting practical deployment.

\parab{Trust Deficit in Content-Based Attacks} 
MAS vulnerabilities are increasingly dominated by content/data manipulation (e.g., prompt infection, psychological manipulation). Addressing these requires reliable tracing of agent interactions and verifiable attribution of actions/data. Existing defenses lack mechanisms to establish such trust, rendering attack-specific mitigations ineffective. For example, defensive strategies against prompt injection \cite{netsafe} cannot reliably determine the origin of malicious content without verifiable interaction histories.

\parab{Inadequate Access Control} 
MAS involve numerous agents with varying capabilities and multi-sourced resources, requiring dynamic, fine-grained access control. Traditional mechanisms (e.g., role-based access control, attribute-based access control) are insufficiently agile to manage real-time interactions or align policies across institutional boundaries and volatile agent interaction contexts. For instance, enforcing data privacy regulations (e.g., GDPR) in cross-institutional or even cross-country MAS remains a significant challenge \cite{achilles-heel}.

These gaps underscore the need for defensive strategies that prioritize verifiable interaction histories, decentralized identity management, and adaptive access control.

\begin{figure*}[ht]
    \centering
    \includegraphics[width=\textwidth]{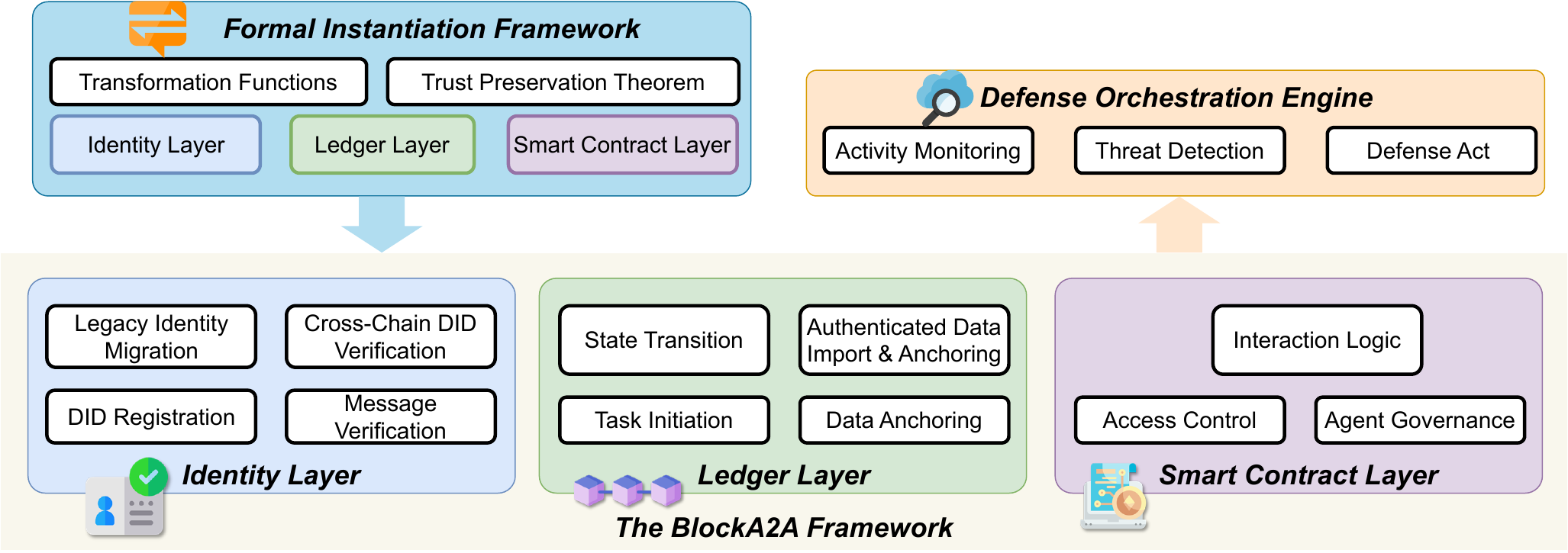}
    \caption{The \sys Framework}
    \label{fig:framework}
\end{figure*}
\section{The \sys Framework}
\label{sec:sys-framework}

In this section, we present \sys, a universal trust framework boosting the transparency, traceability and reliability of agent-to-agent communication. \sys is featured by a three-layer architecture, seeking to mitigate existing security defenses' inherent limitations in identity management, data/interaction anchoring and dynamic access control. 

\subsection{Identity Layer: Decentralized Identity Management}
\label{subsec:identity-layer}

\sys's Identity Layer establishes a trustless, decentralized authentication framework for agent-to-agent interactions. It leverages Decentralized Identifiers (DIDs) and blockchain immutability to ensure secure, scalable, and interoperable identity management for agents. Below, we formalize the architecture, data structures, and protocols with precise definitions of on-chain/off-chain components and processing steps.

\subsubsection{Registration Phase: On-Chain Anchoring and Off-Chain Metadata Storage} \hfill

\parab{On-Chain Data Storage}
The on-chain data storage anchors the existence and integrity of the DID, while enabling tamper-proof verification of the off-chain DID Document. In particular, it includes:

\begin{itemize}
  \item DID Identifier: A unique URI formatted as $$\textsf{did:blocka2a:<algorithm-specific-suffix>},$$registered immutably on the blockchain.
  \item DID Document Hash: A cryptographic hash (e.g., SHA-256) of the off-chain DID Document.
  \item Registration Timestamp: The block number and timestamp of DID registration.
  \item Revocation Status: Boolean flag indicating validity.
\end{itemize}

\parab{Off-Chain Data Storage}
The DID document, a JSON document stored locally by the agent or in a decentralized storage network (e.g., IPFS), contains:

\begin{itemize}
  \item DID Identifier: Consistent with the blockchain-registered DID.
  \item \textsf{publicKey}: Cryptographic public key(s) for signature verification.
  \item \textsf{service}: Network endpoints for inter-agent communication.
  \item \textsf{capabilities}: Machine-readable metadata describing the agent's functionalities/permissions.
  \item \textsf{policy-constraints}: Rules governing interaction, including allowed time, priority and data limit.
  \item \textsf{proof}: (Optional) A self-signed cryptographic proof of the document's integrity.
\end{itemize}

\begin{figure}[htbp]
\centering
\begin{adjustbox}{center,max width=0.8\textwidth}
\begin{lstlisting}[language=json]
{
  "id": "did:blocka2a:ef24a",
  "publicKey": [
    {
      "id": "did:blocka2a:ef24a#key-1",
      "type": "Ed25519VerificationKey2020",
      "publicKeyMultibase": "z6Mk...<base58-encoded-key>"
    }
  ],
  "service": [
    {
      "type": "AgentCommunicationEndpoint",
      "serviceEndpoint": "https://agent-b.example.com/api"
    }
  ],
  "capabilities": {
    "supportedModels": ["GPT-4", "StableDiffusion-v2"],
    "maxComputeTime": "5s",
    "permissions": ["read", "write"]
  },
  "policy-constraints": {
    "allowed_interaction_hours": "09:00-18:00 UTC",
    "max_data_size": "10MB"
  },
  "proof": {
    "type": "Ed25519Signature2020",
    "created": "2023-10-05T12:00:00Z",
    "verificationMethod": "did:blocka2a:ef24a#key-1",
    "proofValue": "z4X2...<base58-encoded-signature>"
  }
}
\end{lstlisting}
\end{adjustbox}
\caption{An example of off-chain DID document.}
\label{DID-document}
\end{figure}

An example of the DID document is shown in Listing~\ref{DID-document}.

\begin{Protocol}[]
    \begin{mdframed}[style=ProtocolFrame, font=\small]
        \centering
        \textbf{Registration Protocol}
        
        \vspace{0.1cm}
        \hrule
        \vspace{0.1cm}
        
        \textit{Input.} An agent $A$ with computational capabilities and access to a blockchain network.
        
        \textit{Output.} A verifiable binding between the agent's identity and its off-chain metadata, anchored on chain.
        
        \vspace{0.1cm}
        \hrule
        \vspace{0.1cm}
        
        \textbf{Protocol Phases and Interaction Flow:}
        
        \begin{center}
            \scalebox{0.9}{
            \begin{tikzpicture}[
                actor/.style={rectangle, draw, text width=2.5cm, align=center, minimum height=1cm},
                message/.style={->, thick},
                process/.style={rectangle, draw, fill=blue!10, text width=3cm, align=center, rounded corners},
                node distance=1.2cm
            ]
                \node (agent) [actor] {Agent $A$};
                \node (contract) [actor, right=2cm of agent] {DID Registry Contract};
                
                \draw[thick, dashed] (agent) -- ++(0,-5.5cm);
                \draw[thick, dashed] (contract) -- ++(0,-5.5cm);
                
                \node (step1) [process, below=0.5cm of agent] {Generate key pair \\ $(sk, pk)$};
                \draw[message] (agent) -- (step1) node[midway, left] {1};

                \node (step2) [process, below=0.5cm of step1] {Generate DID: \\ $\text{DID} = \text{genID}()$};
                \draw[message] (step1) -- (step2) node[midway, right] {2};
                
                \node (step3) [process, below=0.5cm of step2] {Construct DID Document $D$ and its hash $H(D)$};
                \draw[message] (step2) -- (step3) node[midway, left] {3};
                
                \node (tx) [process, below=0.5cm of contract] {Transaction: \\ $\{DID, H(D)\}$};
                \draw[message, bend left=20] (step3.east) -- ++(0.5,0) |- node[pos=0.3, left] {4} (tx.west);
                
                \node (step4) [process, below=0.5cm of tx] {Verify DID: \\ $\text{DID} = \text{isLegitimate}(DID)$};
                \draw[message] (tx) -- (step4) node[midway, right] {5};
                
                \node (step5) [process, below=0.5cm of step4] {On-Chain Record: \\ $\{\text{DID}, H(D), \ldots\}$};
                \draw[message] (step4) -- (step5) node[midway, right] {6};
                
                \draw[message, bend left=25] (step5.south) |- node[pos=0.7, below] {7} ([yshift = -0.7cm] agent |- step5);
            \end{tikzpicture}}
            \end{center} 
        
        \vspace{0.1cm}
        \textbf{Detailed Steps:}
        
        \begin{enumerate}[label=\textbf{\arabic*}, leftmargin=2.5ex, itemsep=0.5ex]
            \item \textbf{Agent Initialization}
                \begin{itemize}
                    \item Agent $A$ generates cryptographic keys:
                        \begin{itemize}
                            \item Private key ($sk$) kept secret;
                            \item Public key ($pk$) for verification.
                        \end{itemize}
                \end{itemize}

            \item \textbf{DID Generation}
                \begin{itemize}
                \item Agent $A$ generates a legitimate DID using its DID controller:
                    $$
                    DID = genID()
                    $$
            \end{itemize}
            
            \item \textbf{DID Document Construction}
                \begin{itemize}
                    \item Agent $A$ constructs DID document $D$ and computes $H(D)$:
                        $$
                        D = \left\{\begin{aligned}
                            & DID, pk, \\
                            & \text{service\_endpoints}, \\
                            & \text{capabilities}, \\
                            & \text{authentication\_methods}, \\
                            & \text{metadata}
                        \end{aligned}\right\}.
                        $$
                \end{itemize}
            
            \item \textbf{Transaction Submission}
                \begin{itemize}
                    \item Agent $A$ submits to the DID Registry Smart Contract with $(DID, H(D))$.
                \end{itemize}
            
            \item \textbf{Smart Contract Processing}
                \begin{itemize}
                    \item The contract verify the legitimacy of the DID string and records:
                        $$
                        \text{On-Chain Record} = \left\{\begin{aligned}
                            & \text{DID}, \text{version}, H(D), \text{timestamp}, \\
                            & \text{revocation\_status}, \text{address}
                        \end{aligned}\right\}.
                        $$
                \end{itemize}
            
            \item \textbf{Completion}
                \begin{itemize}
                    \item Blockchain-anchored identity established.
                \end{itemize}
        \end{enumerate}
    \end{mdframed}
    \caption{The Registration Protocol}
    \label{prot:registration}
\end{Protocol}

\parab{The Registration Protocol}
The specification of our Registration Protocol is provided in Protocol~\ref{prot:registration}. At its core, the protocol leverages cryptographic primitives and smart contract automation to create a tamper-evident link between an agent's identity and its associated attributes. The process begins with the agent generating a key pair $(sk, pk)$, where the private key (\(sk\)) remains secret for authentication, and the public key (\(pk\)) is embedded in a DID (Decentralized Identity) Document (\(D\)). This document encapsulates critical identity components: the DID string, public verification keys, service endpoints for interaction, operational capabilities, authentication methods, and custom metadata. The DID string is generated via a DID controller to ensure its legitimacy. To ensure data integrity, the agent submits a cryptographic hash of the DID document (\(H(D)\)) alongside its DID to the DID Registry Smart Contract. The contract processes this submission by verifying the legitimacy (in both format and uniqueness) of the DID and recording an immutable on-chain entry that includes the DID, hash of the document, timestamp, and revocation status. This workflow ensures that the agent's identity is both cryptographically verifiable and permanently anchored on the blockchain, enabling trustless validation of identity claims across decentralized applications. The protocol's design prioritizes privacy by storing only hashed metadata on-chain, while allowing full DID documents to be retrieved and validated off-chain as needed, thus balancing transparency with data confidentiality. The overall DID document creation and on-chain anchoring process adheres to DID standards like W3C~\cite{w3c-did} to ensure interoperability, security, privacy, and user control over digital identities.

\begin{Protocol}[]
    \begin{mdframed}[style=ProtocolFrame, font=\small]
        \centering
        \textbf{Message Verification Protocol}
        
        \vspace{0.1cm}
        \hrule
        \vspace{0.1cm}
        
        \textit{Input.} Agent $A$ provides a signed message $\sigma = \text{Sign}_{sk_A}(m)$ and its DID string. Agent $B$ has access to the DID Registry Smart Contract and IPFS.
        
        \textit{Output.} Agent $B$ either accepts or rejects Agent $A$'s request based on verification results.
        
        \vspace{0.1cm}
        \hrule
        \vspace{0.1cm}
        
        \textbf{Protocol Phases and Interaction Flow:}
        
        \begin{center}
            \scalebox{0.8}{
            \begin{tikzpicture}[
                actor/.style={rectangle, draw, text width=2cm, align=center, minimum height=1.5cm},
                message/.style={->, thick},
                process/.style={rectangle, draw, fill=blue!10, text width=2cm, align=center, rounded corners},
                node distance=1.2cm
            ]
                \node (agentA) [actor] {Agent $A$};
                \node (agentB) [actor, right=1cm of agentA] {Agent $B$};
                \node (contract) [actor, right=1cm of agentB] {DID Registry Contract};
                
                \draw[thick, dashed] (agentA) -- ++(0,-8cm);
                \draw[thick, dashed] (agentB) -- ++(0,-8cm);
                \draw[thick, dashed] (contract) -- ++(0,-8cm);
                
                \node (step1) [process, below=0.5cm of agentA] {Sign message: \\ $\sigma = \text{Sign}_{sk_A}(m)$};
                \draw[message] (agentA) -- (step1) node[midway, left] {1};
                
                \node (step2) [process, below=0.5cm of step1] {Prepare request: \\ $\{\sigma, \text{DID}_A\}$};
                \draw[message] (step1) -- (step2) node[midway, left] {2};
                
                \ 
                \node (step3) [process, below=0.5cm of agentB] {Receive request \\ from $A$};
                \draw[message, bend left=20] (step2.east) -- ++(0.5,0) |- node[pos=0.3, left] {3} (step3.west);
                
                \node (step4) [process, below=0.5cm of step3] {Query contract: \\ $\text{DID}_A \rightarrow H(D_A)$};
                \draw[message] (step3) -- (step4) node[midway, left] {4};
                
                \node (contractStep) [process, below=0.5cm of contract] {Return: \\ $H(D_A), \text{status}$};
                \draw[message, bend left=20] (step4.east) -- ++(0.5,0) |- node[pos=0.3, left] {5} (contractStep.west);
                
                \node (step5) [process, below=0.5cm of step4] {Fetch $D_A$ from IPFS \\ using $(DID, H(D_A))$};
                \draw[message, bend left=20] (contractStep.south) |- node[pos=0.3, left] {6} (step5.east);
                
                \node (step6) [process, below=0.5cm of step5] {Perform checks: \\ i-iii};
                \draw[message] (step5) -- (step6) node[midway, left] {7};
                
                \node (step7) [process, below=0.5cm of step6] {Accept/Reject \\ request};
                \draw[message] (step6) -- (step7) node[midway, left] {8};
                
                \draw[message, bend left=25] (step7.west) -- node[midway, below] {9} (agentA |- step7);
            \end{tikzpicture}}
        \end{center} 
        
        \vspace{0.2cm}
        \textbf{Detailed Steps:}
        
        \begin{enumerate}[label=\textbf{\arabic*}, leftmargin=2.5ex, itemsep=0.5ex]
            \item \textbf{Message Signing}
                \begin{itemize}
                    \item Agent $A$ initiates a request to Agent $B$ by signing payload $m$ with its private key:
                        \[
                        \sigma = \text{Sign}_{sk_A}(m)
                        \]
                    \item The request package includes:
                        \begin{itemize}
                            \item The signed message $m || \sigma$.
                            \item Agent $A$'s DID string (e.g., $\text{did}:\text{blocka2a}:\text{eth}:\text{abc123}$).
                        \end{itemize}
                \end{itemize}
            
            \item \textbf{On-Chain Verification}
                \begin{itemize}
                    \item Agent $B$ queries the DID Registry Smart Contract using $A$'s DID to retrieve:
                        \begin{itemize}
                            \item $H(D_A)$: Cryptographic hash of Agent $A$'s DID document.
                            \item $\text{revocation\_status}$: To ensure the DID is active.
                        \end{itemize}
                \end{itemize}
            
            \item \textbf{Off-Chain Validation}
                \begin{itemize}
                    \item Agent $B$ fetches $D_A$ from IPFS using $H(D_A)$ and performs:
                        \begin{enumerate}[label=\roman*.]
                            \item \textbf{Integrity Check}: Compute $H(D_A')$ and compare with the on-chain $H(D_A)$.
                            \item \textbf{Ownership Verification}: Verify $\sigma$ using $pk_A$ from $D_A$.
                            \item \textbf{Permissions Check}: Validate $A$'s request against $D_A.\text{capabilities}$.
                        \end{enumerate}
                \end{itemize}
            
            \item \textbf{Decision Making}
                \begin{itemize}
                    \item Agent $B$ accepts the request if all checks pass:
                        \[
                        \text{Accept} \iff \text{Integrity} \land \text{Ownership} \land \text{Permissions}
                        \]
                    \item Otherwise, the request is rejected with appropriate error messaging.
                \end{itemize}
        \end{enumerate}
    \end{mdframed}
    \caption{The Message Verification Protocol}
    \label{prot:message-verification}
\end{Protocol}

\subsubsection{Verification Phase: Cryptographic Identity Validation}

Building upon the foundations established by the Registration Protocol, the Message Verification Protocol (specified in Protocol~\ref{prot:message-verification}) enables secure, decentralized authentication and authorization of interactions between agents within the blockchain ecosystem. This protocol leverages the cryptographically anchored identities created during registration to validate messages without relying on centralized authorities, establishing an application-layer message trust over traditional TLS-based authentications. In particular, when Agent \(A\) initiates a request to Agent \(B\), it signs the message payload (\(m\)) using its private key (\(sk_A\)), generating a digital signature (\(\sigma\)). This signed message, accompanied by Agent \(A\)'s DID, is transmitted to Agent \(B\), who orchestrates a multi-step verification process. First, Agent \(B\) queries the DID Registry Smart Contract—using the provided DID—to retrieve the cryptographic hash of Agent \(A\)'s DID document (\(H(D_A)\)) and its revocation status, ensuring the identity remains active. Next, Agent \(B\) fetches the full DID document (\(D_A\)) from IPFS using \( (DID, H(D_A)) \) and performs three critical checks: (i) verifying the document's integrity by recomparing its hash with the on-chain record, (ii) authenticating the signature using the public key (\(pk_A\)) embedded in \(D_A\), and (iii) validating that Agent \(A\) possesses the necessary permissions to execute the requested action. The protocol ensures that only agents with valid, unrevoked identities and appropriate capabilities can interact, thereby enforcing trust boundaries defined during registration. By linking verification directly to the immutable on-chain DID records established in the Registration Protocol, this mechanism creates a robust framework for verifiable, permissioned communication in decentralized systems. The modular design allows the protocol to be integrated into various applications, from secure data sharing to automated service provisioning, while maintaining end-to-end cryptographic security.

Example: If Agent A requests to update supply chain data from Agent B: Agent B validates Agent A's DID, confirms its \textsf{write} permission, and accepts the request only if all checks pass.

\begin{Protocol}[t]
    \begin{mdframed}[style=ProtocolFrame, font=\small]
        \centering
        \textbf{Cross-Chain DID Validation Protocol}
        
        \vspace{0.1cm}
        \hrule
        \vspace{0.1cm}
        
        \textit{Input.} Agent $A$ on Chain $X$ provides $H(D_A)$ and DID document $D_A$. Agent $B$ on Chain $Y$ has access to Chain $Y$'s Federated Trust Anchor Contract.
        
        \textit{Output.} Agent $B$ verifies the authenticity of $D_A$ from Chain $X$ via cross-chain interoperability.
        
        \vspace{0.1cm}
        \hrule
        \vspace{0.1cm}
        
        \textbf{Protocol Phases and Interaction Flow:}
        
        \begin{center}
            \scalebox{0.85}{
            \begin{tikzpicture}[
                actor/.style={rectangle, draw, text width=3cm, align=center, minimum height=1.2cm},
                message/.style={->, thick},
                process/.style={rectangle, draw, fill=green!10, text width=3cm, align=center, rounded corners},
                node distance=1cm
            ]
                \node (agentA) [actor] {Agent $A$ \\ (Chain $X$)};
                \node (bridge) [actor, right=1cm of agentA] {Cross-Chain Bridge};
                \node (agentB) [actor, right=1cm of bridge] {Agent $B$ \\ (Chain $Y$)};
                
                \node (step1) [process, below=0.8cm of agentA] {Submit $H(D_A)$ \\ to bridge};
                \node (step2) [process, below=0.8cm of bridge] {Relay $H(D_A)$ \\ to Chain $Y$};
                \node (step3) [process, below=0.8cm of agentB] {Query \\ Trust Anchor};
                
                \draw[message] (agentA) -- (step1);
                \draw[message] (step1.east) -- node[midway, above] {1} (step2.west |- step1);
                \draw[message] (bridge) -- (step2);
                \draw[message] (step2.east) -- node[midway, above] {2} (step3.west |- step2);
                \draw[message] (agentB) -- (step3);
            \end{tikzpicture}}
        \end{center}

        \vspace{0.2cm}
        \textbf{Detailed Steps:}
        
        \begin{enumerate}[label=\textbf{\arabic*}, leftmargin=2.5ex, itemsep=0.5ex]
            \item \textbf{Cross-Chain Hash Relay}
                \begin{itemize}
                    \item \textbf{Bridge Interaction:}
                        \begin{itemize}
                            \item Agent $A$ submits hash $H(D_A)$ through cross-chain bridge.
                            \item Bridge validates Chain $X$ consensus proof for $H(D_A)$.
                        \end{itemize}
                    \item \textbf{Trust Anchor Update:}
                        \begin{itemize}
                            \item Relayed hash stored in Chain $Y$'s contract: 
                            $$\text{AnchorStore} = \left\{\begin{aligned}
                                & H(D_A), \\
                                & \text{source\_chain}, \\
                                & \text{block\_number}
                            \end{aligned}\right\}.$$
                        \end{itemize}
                \end{itemize}
            
            \item \textbf{Federated Verification}
                \begin{itemize}
                    \item \textbf{Document Retrieval:}
                        \begin{itemize}
                            \item Agent $B$ requests $D_A$ from Agent $A$ via off-chain channel.
                            \item Agent $A$ provides $D_A$ with cryptographic signature.
                        \end{itemize}
                    \item \textbf{Consensus Validation:}
                        \begin{itemize}
                            \item Agent $B$ verifies: 
                            $$\text{Valid} \iff \left\{\begin{aligned}
                                & \texttt{verifySig}(D_A, pk_A) \\
                                & \land \; H(D_A) \equiv \text{AnchorStore} \\
                                & \land \; \text{timestamp} \leq \tau_{\text{max}}
                            \end{aligned}\right\}$$
                        \end{itemize}
                \end{itemize}
        \end{enumerate}
    \end{mdframed}
    \caption{The Cross-Chain DID Validation Protocol}
    \label{prot:cross-chain-validation}
\end{Protocol}

\subsubsection{Cross-Domain Interoperability}

Building on the single-chain identity infrastructure established by the Registration Protocol and the intra-chain verification mechanisms of the Message Verification Protocol, the Cross-Chain DID Validation Protocol addresses the challenge of trust propagation across heterogeneous blockchain networks. This protocol, defined in Protocol~\ref{prot:cross-chain-validation}, extends the cryptographic anchoring of DIDs (established in the Registration Protocol) to enable interoperable identity verification between agents on separate chains (e.g., Agent \(A\) on Chain \(X\) and Agent \(B\) on Chain \(Y\)). Leveraging cross-chain interoperability protocols such as IBC (for Cosmos SDK chains)~\cite{ibc}, Chainlink CCIP (for EVM chains)~\cite{ccip} or HyperService~\cite{hyperservice}, the protocol facilitates the secure relay of DID document hashes (\(H(D_A)\)) between chains via a trustless bridge mechanism.

The workflow begins with Agent \(A\) submitting \(H(D_A)\) to the cross-chain bridge, which validates the hash against Chain \(X\)'s consensus proof before relaying it to Chain \(Y\)'s Federated Trust Anchor Contract. This contract stores the hash alongside metadata (e.g., source chain, block number) to create a cross-chain trust anchor. When Agent \(B\) receives \(D_A\) via an off-chain channel (accompanied by Agent \(A\)'s cryptographic signature), it verifies three critical properties: \first the signature on \(D_A\) confirms Agent \(A\)'s ownership of the associated private key (consistent with the Registration Protocol's key-pair generation); \second the hash of \(D_A\) matches the anchored \(H(D_A)\) on Chain \(Y\), ensuring document integrity; and \third the timestamp of the anchor record falls within an acceptable validity window (\(\tau_{\text{max}}\)), mitigating replay attacks.

By extending the single-chain identity anchors from the Registration Protocol into a cross-chain trust framework, this protocol enables seamless validation of decentralized identities across heterogeneous ecosystems. MAS It resolves the “identity silo” problem inherent in single-chain systems, allowing agents to leverage their registered identities for cross-chain interactions (e.g., asset transfers, service invocations) while maintaining the cryptographic security and decentralization principles established in prior protocols. The integration of cross-chain bridges and federated trust anchors ensures that trust is neither centralized nor reliant on pre-established inter-chain alliances, thus enabling a scalable, permissionless framework for multi-chain identity interoperability.

\begin{Protocol}[]
    \begin{mdframed}[style=ProtocolFrame, font=\small]
        \centering
        \textbf{Legacy Identity to DID Migration Protocol - Part 1}
        
        \vspace{0.1cm}
        \hrule
        \vspace{0.1cm}
        
        \textit{Input.} AI agent or managing entity with a legacy identity system credential; access to the Migration committee (for decentralized validation) and Oracle services (for legacy system bridging). \\
        \textit{Output.} A Sybil-resistant DID-based master credential ($MC$) bound to the agent, with optional decommissioning of the legacy identity.
        
        \vspace{0.1cm}
        \hrule
        \vspace{0.1cm}
        
        \textbf{Protocol Phases and Interaction Flow:}
        
        \begin{center}
            \scalebox{0.75}{
            \begin{tikzpicture}[
                actor/.style={rectangle, draw, text width=3.5cm, align=center, minimum height=1.2cm, font=\small},
                message/.style={->, thick, font=\small},
                process/.style={rectangle, draw, fill=green!10, text width=3.5cm, align=center, rounded corners, font=\small},
                node distance=1cm
            ]
                \node (agent) [actor] {AI Agent / \\ Managing Entity};
                \node (Oracle) [actor, right=1.5cm of agent] {Oracle Service};
                \node (committee) [actor, right=1.5cm of Oracle] {Migration Committee};

                \draw[thick, dashed] (agent) -- ++(0,-8cm);
                \draw[thick, dashed] (Oracle) -- ++(0,-8cm);
                \draw[thick, dashed] (committee) -- ++(0,-8cm);
                
                \node (step1) [process, below=0.5cm of agent] {1. Declare migration intent; \\ Provide legacy ID};
                \node (step2) [process] at (Oracle |- step1) {2. Verify legacy identity \\ via Oracle};
                \node (step3) [process] at (committee |- step2) {3. Provisionally map \\ DID $\leftrightarrow$ LegacyID};
                \node (step4) [process, below=0.5cm of step3] {4. MPC-based Sybil check \\ (duplication prevention)};
                \node (step5) [process, below=0.5cm of step4] {5. Issue master credential \\ $MC$ with Sybil proof};
                \node (step6) [process] at ([yshift=-1.5cm, xshift=1.8cm] agent |- step5) {6. (Optional) Generate \\ context-specific credentials};
                
                \draw[message] (agent) -- (step1);
                \draw[message] (step1.east) -- node[midway, above] {LegacyID} node[midway, below] {intent} (step2.west);
                \draw[message] (Oracle) -- (step2);
                \draw[message] (step2.east) -- node[midway, above] {Attestation} (step3.west);
                \draw[message] (committee) -- (step3);
                \draw[message] (step3) -- (step4);
                \draw[message] (step4) -- (step5);
                \draw[message] (step5.west) -- node[midway, above] {$MC$} (agent |- step5);
                \draw[message] (step5.south) |- (step6.east);
                \draw[message] (agent |- step1.south) |- (step6.west);
            \end{tikzpicture}}
        \end{center}
    \end{mdframed}
    \caption{Legacy Identity to DID Migration Protocol - Part 1: Overview and Interaction Flow}
    \label{prot:legacy-to-did-migration-part1}
\end{Protocol}

\begin{Protocol}[]
    \begin{mdframed}[style=ProtocolFrame, font=\small]
        \centering
        \textbf{Legacy Identity to DID Migration Protocol - Part 2}
        
        \vspace{0.1cm}
        \hrule
        \vspace{0.1cm}
        
        \textbf{Detailed Steps:}
        
        \begin{enumerate}[label=\textbf{\arabic*}, leftmargin=2.5ex, itemsep=0.5ex]
            \item \textbf{Legacy Identity Declaration and Intent}
                \begin{itemize}
                    \item The agent/managing entity initiates migration by submitting:
                        \begin{itemize}
                            \item A unique legacy system identifier (e.g., UUID, enterprise ID),
                            \item A signed declaration of intent, 
                        \end{itemize}
                        to the Migration committee via Oracle intermediation.
                \end{itemize}
            
            \item \textbf{Oracle-Mediated Legacy Identity Validation}
                \begin{itemize}
                    \item An authorized Oracle retrieves and verifies the legacy identity:
                        \[
                        \text{Attestation} = \left\{
                            \begin{aligned}
                                & \text{LegacyID}, \; \langle \text{Attributes} \rangle, \; \text{Timestamp}, \\
                                & \sigma_{\text{Oracle}} = \text{Sign}_{sk_{\text{Oracle}}}(\text{LegacyID} \| \text{Attributes})
                            \end{aligned}
                        \right\}
                        \]
                    \item It binds legacy attributes to the agent's claimed identity.
                    \item Sensitive attributes are kept in secret form $\langle \text{Attributes} \rangle$.
                \end{itemize}
            
            \item \textbf{DID Generation and Provisional Linking}
                \begin{itemize}
                    \item The agent generates a new DID ($D_{\text{agent}}$) and key pair ($sk_{\text{agent}}, pk_{\text{agent}}$) following the Registration Protocol's cryptographic standards.
                    \item The committee validates the Oracle's attestation ($\sigma_{\text{Oracle}}$) and stores a provisional mapping:
                        \[
                        \text{ProvisionalStore} = \left\{ D_{\text{agent}} \rightarrow (\text{LegacyID}, \langle \text{Attestation} \rangle) \right\}
                        \]
                \end{itemize}
            
            \item \textbf{Privacy-Preserving Sybil Resistance Check}
                \begin{itemize}
                    \item Using secure multi-party computation (MPC), the committee checks for duplicate identities across its registry:
                        \[
                        \text{Duplicate} = \exists D' \neq D_{\text{agent}} \; \text{s.t.} \; f(A(D_{\text{agent}}), A(D')) = \text{true}
                        \]
                        where \(A(D)\) denotes the attribute set of DID \(D\), and \(f\) is a threshold-based similarity function.
                    \item The MPC protocol ensures raw attribute data remains confidential, with a quorum of \(t\) committee members required for validation.
                \end{itemize}
            
            \item \textbf{Master Credential Issuance and Anchoring}
                \begin{itemize}
                    \item Upon passing the Sybil check, the committee issues a master credential:
                        \[
                        MC = \left\{
                            \begin{aligned}
                                & D_{\text{agent}}, \; \text{LegacyID}, \; \text{Attributes}, \\
                                & \text{Issuer}_{\text{Migration}}, \; \text{Timestamp}, \; \sigma_{\text{committee}}, \; \pi_{\text{sybil}}
                            \end{aligned}
                        \right\}
                        \]
                        containing a cryptographic proof of Sybil resistance (\(\pi_{\text{sybil}}\)).
                    \item \(MC\) is anchored to the DID registry, and the agent receives secure access to the credential via its private key.
                \end{itemize}
            
            \item \textbf{Optional Context-Specific Identity Derivation}
                \begin{itemize}
                    \item For granular interactions, the agent generates ephemeral DIDs (\(D_{\text{context}}\)) from \(D_{\text{agent}}\), providing selective disclosures:
                        \[
                        P = \left\{
                            D_{\text{context}}, \; \text{SelectiveDisclosure}(MC), \; \sigma_{\text{agent}}
                        \right\}
                        \]
                    \item This maintains privacy by revealing only context-relevant attributes while proving linkage to the master credential.
                \end{itemize}
            
            \item \textbf{Legacy System Decommissioning (Optional)}
                \begin{itemize}
                    \item The legacy system updates the identity status to one of:
                        \[
                        \text{Status} \in \{\text{Migrated} \rightarrow D_{\text{agent}}, \; \text{Revoked}, \; \text{Archived}\}
                        \]
                    \item Future access requests are redirected to the DID-based verification workflow, as defined in the Message Verification Protocol.
                \end{itemize}
        \end{enumerate}
    \end{mdframed}
    \caption{Legacy Identity to DID Migration Protocol - Part 2: Detailed Steps}
    \label{prot:legacy-to-did-migration-part2}
\end{Protocol}

\subsubsection{Legacy System Migration}

The Legacy Identity to DID Migration Protocol enables secure transition of centralized agent identities (e.g., enterprise UUID, government IDs) into \sys's decentralized identity framework, ensuring compatibility while enhancing security and interoperability. Building on the Registration Protocol's cryptographic foundations, this protocol bridges legacy systems with self-sovereign DIDs through a structured, auditable process. As shown in Protocol~\ref{prot:legacy-to-did-migration-part1} \& \ref{prot:legacy-to-did-migration-part2}, the core workflow consists of:
\begin{itemize}
  \item \textit{Legacy Identity Validation}: Agents first declare migration intent and provide their legacy identifier (e.g., corporate ID) to an Oracle service. The Oracle, built upon techniques from either DECO~\cite{deco} or Town Crier~\cite{town-crier}, verifies the legacy identity's authenticity and issues a signed attestation, linking attributes (e.g., role, permissions) to the claimed identity without viewing raw attribute data. This mirrors the Registration Protocol's trust-by-signature model, ensuring non-repudiation.
  \item \textit{DID Provisioning \& Linkage}: Following the Registration Protocol's standards, agents generate a new DID and key pair. The Migration committee (consisting of blockchain nodes) validates the Oracle's attestation and creates a temporary link between the legacy ID and new DID in its decentralized registry, ensuring compatibility with existing DID document structures and on-chain records.
 \item \textit{Sybil Resistance Check}: The committee checks for duplicate identities across its registry to prevent malicious multiple accounts (Sybil attacks). For sensitive attributes stored in secret form (like secret-shared or encrypted), the checks can be performed privacy-preservingly via secure multi-party computation (MPC). This step ensures each legacy identity maps to exactly one DID without exposing sensitive attributes, extending the security guarantees of prior verification protocols.
  \item \textit{Master Credential Issuance}: A master credential is issued, formally binding the legacy identity to the DID. This credential includes a cryptographic proof of the Sybil check and is anchored in the DID registry, allowing seamless use in decentralized applications. It aligns with the credential schemas used in the Message Verification Protocol, enabling consistent verification across systems.
\end{itemize}

The optional extensions for finer-grained and smooth transition to DID include:
\begin{itemize}
  \item \textit{Context-Specific Identities}: Agents can generate temporary DIDs for specific services or contexts (e.g., HR systems), sharing only necessary attributes to protect privacy while proving identity linkage.
  \item \textit{Legacy System Integration}: Organizations can mark legacy identities as "migrated" and route new access requests to the DID framework, allowing a gradual shift away from centralized systems while maintaining backward compatibility.
Value Proposition
\end{itemize}
This protocol acts as a bridge, allowing organizations to leverage existing identity investments while gaining the benefits of decentralization: cryptographic security, resistance to single points of failure, and interoperability across blockchain networks (as enabled by the Cross-Chain DID Validation Protocol). By formalizing the migration process, it ensures a secure, auditable path for legacy systems to join the decentralized identity ecosystem, without compromising on security or operational continuity.

\subsection{Ledger Layer: Selective On-Chain Provenance for Immutable Accountability}
\label{subsec:ledger-layer}

\sys's Ledger Layer ensures non-repudiation, auditability, and long-term data integrity for collaborative task execution across agents, by strategically anchoring high-value interaction metadata on the blockchain. This layer avoids the prohibitive costs of full on-chain storage while preserving verifiability through cryptographic commitments and multi-party consensus.

\subsubsection{Architecture Overview}

The Ledger Layer operates on a selective provenance model, where only critical interaction data is hashed and recorded on-chain. This minimizes storage overhead while enabling tamper-evident audit trails. Key components include:

\begin{itemize}
  \item Provenance Smart Contract : Manages hash anchoring, multi-signature validation, and dispute resolution.
  \item Off-Chain Data Repository : Stores full interaction payloads (e.g., task details, state changes) in distributed systems like IPFS or Filecoin.
  \item Cryptographic Primitives : SHA-256 for hashing, BLS multi-signatures for consensus, and Merkle trees for batch verification.
\end{itemize}

\parab{Provenance Principle: Minimal On-Chain Storage}
The overall objective of the Ledger Layer is to record \textit{minimally sufficient data} for reconstructing interactions and enforce accountability. This goal encompasses three aspects:

\begin{itemize}
  \item \textit{Non-repudiation}: Cryptographic proof of participation in tasks;
  \item \textit{State Consistency}: Cross-agent consensus on task lifecycle milestones;
  \item \textit{Data Integrity}: Tamper-evident anchoring of off-chain artifacts.
\end{itemize}

On the other hand, the raw input/output files, transient intermediate states, and non-critical metadata remain off-chain, which largely reduces blockchain bloat compared to full-state recording.

\begin{Protocol}[]
    \begin{mdframed}[style=ProtocolFrame, font=\small]
        \centering
        \textbf{Task Initiation Protocol}
        
        \vspace{0.1cm}
        \hrule
        \vspace{0.1cm}
        
        \textit{Input.} 
        \begin{itemize}
            \item \textbf{Task Metadata}: Initiator DID ($\text{DID}_{\text{init}}$), participant DIDs ($\text{DID}_1, \ldots, \text{DID}_n$), task description, deadline.
            \item \textbf{Timestamp}: Blockchain timestamp ($t$).
        \end{itemize}
        
        \textit{Output.} 
        \begin{itemize}
            \item On-chain entry in Provenance Smart Contract: $\{\text{H}_{\text{task}}, t, \text{status: "initiated"}\}$
            \item Off-chain storage of full task metadata
        \end{itemize}
        
        \vspace{0.1cm}
        \hrule
        \vspace{0.1cm}
        
        \textbf{Protocol Phases and Interaction Flow:}
        
        \begin{center}
            \scalebox{0.75}{
            \begin{tikzpicture}[
                actor/.style={rectangle, draw, text width=3.5cm, align=center, minimum height=1.5cm},
                message/.style={->, thick},
                process/.style={rectangle, draw, fill=blue!10, text width=3.5cm, align=center, rounded corners},
                node distance=1.2cm
            ]
                \node (initiator) [actor] {Task Initiator};
                \node (contract) [actor, right=1cm of initiator] {Provenance Contract};
                \node (storage) [actor, right=0.8cm of contract] {Off-chain Storage};
                
                \draw[thick, dashed] (initiator) -- ++(0,-6cm);
                \draw[thick, dashed] (contract) -- ++(0,-6cm);
                \draw[thick, dashed] (storage) -- ++(0,-6cm);
                
                \node (step1) [process, below=0.5cm of initiator] {Assemble task metadata};
                \draw[message] (initiator) -- (step1) node[midway, left] {1};
                
                \node (step2) [process, below=0.5cm of step1] {Compute: \\ $H_{\text{task}} = \text{SHA-256}(DID_{\text{init}} \| DID_1 \|$ $\cdots \| DID_n \| \text{description} \| t)$};
                \draw[message] (step1) -- (step2) node[midway, left] {2};
                
                \node (step3) [process] at (contract.south |- step2) {Record on-chain: \\ $\{H_{\text{task}}, t, \text{status: "initiated"}\}$};
                \draw[message, bend left=20] (step2.east) -- ++(0.2,0) |- node[midway, above] {3} (step3.west);
                
                \node (step4) [process, below=4cm of storage] {Store full metadata};
                \draw[message, bend left=20] (initiator |- step4.west) -- node[midway, above] {4} (step4.west);
                
                \draw[message] (step3.south) |- node[pos=0.7, below] {5} ([yshift=-0.5cm] initiator |- step2.south);
                \draw[message, bend left=25] (step4.south) |- node[pos=0.7, below] {6} ([yshift=-0.5cm] initiator |- step4);
            \end{tikzpicture}}
        \end{center} 
        
        \vspace{0.2cm}
        \textbf{Detailed Steps:}
        
        \begin{enumerate}[label=\textbf{\arabic*}, leftmargin=2.5ex, itemsep=0.5ex]
            \item \textbf{Metadata Collection}
                \begin{itemize}
                    \item The task initiator assembles all required metadata including:
                    \begin{itemize}
                        \item Their own DID ($\text{DID}_{\text{init}}$);
                        \item List of participant DIDs ($\text{DID}_1, \ldots, \text{DID}_n$);
                        \item Detailed task description;
                        \item Deadline timestamp.
                    \end{itemize}
                \end{itemize}
            
            \item \textbf{Hash Computation}
                \begin{itemize}
                    \item Compute the cryptographic hash of the task parameters:
                    \[
                    H_{\text{task}} = \text{SHA-256}(DID_{\text{init}} \| DID_1 \| \cdots \| DID_n \| \text{description} \| t),
                    \]
                    where $\|$ denotes concatenation of the components.
                \end{itemize}
            
            \item \textbf{On-chain Recording}
                \begin{itemize}
                    \item The initiater submit $H_{\text{task}}$ to the Provenance Smart Contract;
                    \item Contract records the following on-chain entry:
                    \[
                    \text{On-chain Entry} = \{H_{\text{task}}, t, \text{status: "initiated"}\}.
                    \]
                \end{itemize}
            
            \item \textbf{Off-chain Storage}
                \begin{itemize}
                    \item The full task metadata (including descriptions and participant details) is stored in an off-chain repository.
                    \item Link between on-chain hash and off-chain data is established via $H_{\text{task}}$.
                \end{itemize}
            
            \item \textbf{Confirmation}
                \begin{itemize}
                    \item Provenance Contract returns transaction receipt to initiator.
                    \item Off-chain storage system confirms successful metadata storage.
                \end{itemize}
        \end{enumerate}
    \end{mdframed}
    \caption{The Task Initiation Protocol}
    \label{prot:task-initiation}
\end{Protocol}

\subsubsection{Task Initiation}
The \textit{Task Initiation Protocol} establishes a structured framework for initiating multi-agent tasks with verifiable provenance, integrating on-chain immutability and off-chain efficiency. As depicted in Protocol \ref{prot:task-initiation}, the protocol begins with the task initiator assembling critical metadata, including decentralized identifiers (DIDs) for both the initiator and participants, a detailed task description, and a deadline, paired with a blockchain timestamp (\(t\)). A cryptographic hash (\(H_{\text{task}}\)) is then computed using SHA-256 to uniquely represent the concatenated task parameters, ensuring data integrity through cryptographic binding. This hash, alongside the timestamp and an "initiated" status flag, is recorded on-chain via the Provenance Smart Contract, creating an immutable audit trail. Concurrently, the full task metadata—too voluminous for efficient on-chain storage—is securely stored off-chain, with a cryptographic link maintained through \(H_{\text{task}}\) to enable seamless cross-referencing between the compact on-chain record and detailed off-chain data. The protocol ensures atomicity through a two-fold confirmation mechanism: the smart contract returns a transaction receipt to acknowledge on-chain recording, while the off-chain storage system validates successful metadata deposition. This design balances blockchain's tamper-resistance with practical data management, providing a foundation for traceable task lifecycle management in decentralized ecosystems.

\begin{Protocol}[]
    \begin{mdframed}[style=ProtocolFrame, font=\small]
        \centering
        \textbf{State Transition Validation Protocol}
        
        \vspace{0.1cm}
        \hrule
        \vspace{0.1cm}
        
        \textit{Trigger.} Completion of a task milestone \\ (e.g., delivery confirmation, QA approval).
        
        \textit{Input.} 
        \begin{itemize}
            \item Task hash: $H_{\text{task}}$
            \item Milestone identifier: "milestone-X"
            \item BLS signatures from involved agents: $\sigma_1, \ldots, \sigma_n$
        \end{itemize}
        
        \textit{Output.} 
        \begin{itemize}
            \item Updated on-chain status: \\ $\text{On-Chain Entry}[H_{\text{task}}].\text{status} = \text{"milestone-X}_{\text{verified}}"$
            \item Transaction receipt from smart contract
        \end{itemize}
        
        \vspace{0.1cm}
        \hrule
        \vspace{0.1cm}
        
        \textbf{Protocol Phases and Interaction Flow:}
        
        \begin{center}
            \scalebox{0.75}{
            \begin{tikzpicture}[
                actor/.style={rectangle, draw, text width=4cm, align=center, minimum height=1cm},
                message/.style={->, thick},
                process/.style={rectangle, draw, fill=blue!10, text width=4cm, align=center, rounded corners},
                node distance=1.2cm
            ]
                \node (agents) [actor] {Involved Agents};
                \node (aggregator) [actor, right=1cm of agents] {Signature Aggregator};
                \node (contract) [actor, right=1cm of aggregator] {Provenance Contract};
                
                \draw[thick, dashed] (agents) -- ++(0,-5.5cm);
                \draw[thick, dashed] (aggregator) -- ++(0,-5.5cm);
                \draw[thick, dashed] (contract) -- ++(0,-5.5cm);
                
                \node (step1) [process, below=0.5cm of agents] {Sign: \\ $\sigma_i = \text{BLS.Sign}_{sk_i}(H_{\text{task}} \|$ $\text{"milestone-X"})$};
                \draw[message] (agents) -- (step1) node[midway, left] {1};
                
                \node (step2) [process] at (aggregator |- step1) {Collect signatures \\ $\sigma_1, \ldots, \sigma_n$};
                \draw[message, bend left=20] (step1.east) -- node[midway, above] {2} (step2.west);
                
                \node (step3) [process, below=0.5cm of step2] {Compute: \\ $\Sigma = \text{BLS.Aggregate}(\sigma_1, \ldots, \sigma_n)$};
                \draw[message] (step2) -- (step3) node[midway, left] {3};
                
                \node (step4) [process] at (contract |- step3) {Verify $\Sigma$ against public keys};
                \draw[message, bend left=20] (step3.east) -- node[midway, above] {4} (step4.west);
                
                \node (step5) [process, below=0.5cm of step4] {Update on-chain status: \\ $\text{status} = \text{"milestone-X}_{\text{verified}}"$};
                \draw[message] (step4) -- (step5) node[midway, right] {5};
                
                \draw[message, bend left=25] (step5.west) -- node[midway, below] {6} (aggregator |- step5);
            \end{tikzpicture}}
        \end{center} 
        
        \vspace{0cm}
        \textbf{Detailed Steps:}
        
        \begin{enumerate}[label=\textbf{\arabic*}, leftmargin=2.5ex, itemsep=0.5ex]
            \item \textbf{Multi-Signature Collection}
                \begin{itemize}
                    \item Each involved agent $i$ generates a BLS signature over the task hash and milestone identifier:
                    \[
                    \sigma_i = \text{BLS.Sign}_{sk_i}(H_{\text{task}} \| \text{"milestone-X"}) \quad \text{for } i = 1, \ldots, n
                    \]
                    where $sk_i$ is the agent's private key.
                \end{itemize}
            
            \item \textbf{Signature Aggregation}
                \begin{itemize}
                    \item A designated aggregator collects all individual signatures $\sigma_1, \ldots, \sigma_n$.
                    \item The aggregator computes the BLS aggregate signature:
                    \[
                    \Sigma = \text{BLS.Aggregate}(\sigma_1, \ldots, \sigma_n).
                    \]
                \end{itemize}
            
            \item \textbf{On-Chain Submission}
                \begin{itemize}
                    \item The aggregator submits $(H_{\text{task}}, \Sigma, \text{"milestone-X"}, \{DID_i\})$ to the Provenance Contract, where $\{DID_i\}$ is the set of agent DIDs.
                    \item This triggers the contract's verification process.
                \end{itemize}
            
            \item \textbf{Signature Verification}
                \begin{itemize}
                    \item The contract retrieves the agents' public keys from IPFS via Oracle.
                    \item It verifies the aggregate signature $\Sigma$ against these public keys:
                    \[
                    \text{Verify}(\Sigma, H_{\text{task}} \| \text{"milestone-X"}, pk_1, \ldots, pk_n).
                    \]
                \end{itemize}
            
            \item \textbf{State Update}
                \begin{itemize}
                    \item If the verification succeeds, the contract updates the on-chain entry:
                    \[
                    \text{On-Chain Entry}[H_{\text{task}}].\text{status} = \text{"milestone-X}_{\text{verified}}."\
                    \]
                    \item The contract emits an event notifying the state change.
                \end{itemize}
            
            \item \textbf{Confirmation}
                \begin{itemize}
                    \item The smart contract returns a transaction receipt to the aggregator.
                    \item The receipt includes the new status and transaction ID.
                \end{itemize}
        \end{enumerate}
    \end{mdframed}
    \caption{The State Transition Validation Protocol}
    \label{prot:state-validation}
\end{Protocol}

\subsubsection{Task State Transition}
The State Transition Validation Protocol (Protocol \ref{prot:state-validation}) introduces a cryptographic mechanism for verifying task milestone completions in decentralized ecosystems, ensuring tamper-resistant state transitions through BLS multi-signature aggregation and on-chain validation. Triggered by events like deliverable confirmation or quality assurance approval, this protocol operates by first requiring involved agents to generate individual BLS signatures (\(\sigma_i\)) over the task hash (\(H_{\text{task}}\)) and a specific milestone identifier (e.g., "milestone-X"). These signatures serve as cryptographic endorsements of the milestone achievement, each bound to an agent's private key (\(sk_i\)) to ensure non-repudiation. A designated aggregator then consolidates these individual signatures into a single aggregate signature (\(\Sigma\)) using BLS aggregation—a process that reduces \(n\) distinct signatures into a constant-size proof, minimizing communication overhead while preserving collective accountability.

Upon aggregation, the tuple \((H_{\text{task}}, \Sigma, \text{"milestone-X"}, \{DID_i\})\) is submitted to the Provenance Smart Contract, which initiates a verification routine. The contract retrieves participants' public keys (\(pk_1, \ldots, pk_n\)) from the IPFS via Oracle, and cross-checks the aggregate signature against the combined public key set, ensuring that all required stakeholders have consented to the milestone transition. If validation succeeds, the contract updates the on-chain status of \(H_{\text{task}}\) to "milestone-X\(_{\text{verified}}\)," immutably recording the state change and emitting an event to notify the ecosystem. This design capitalizes on BLS cryptography's efficiency, enabling scalable consensus without compromising verifiability: the aggregate signature maintains the same size regardless of participant count, making it ideal for large-scale decentralized workflows.

By mandating multi-signature consensus for state updates, the protocol ensures that transitions occur only with the explicit agreement of all relevant agents, fortifying trust in distributed task lifecycles. The separation of off-chain signature aggregation—where computational intensity is managed efficiently—and on-chain verification—where finality is guaranteed by blockchain immutability—strikes a balance between performance and security. This architecture provides a robust foundation for tracking task progression in multi-agent collaboration, enabling auditable, consensus-driven state transitions without sacrificing scalability.

\begin{Protocol}[]
    \begin{mdframed}[style=ProtocolFrame, font=\small]
        \centering
        \textbf{Data Anchoring Protocol}
        
        \vspace{0.1cm}
        \hrule
        \vspace{0.1cm}
        
        \textit{Use Case.} Long-term integrity verification of large payloads (e.g., contracts, audit logs).
        
        \textit{Input.} 
        \begin{itemize}
            \item Data payload: $P$
            \item Timestamp: $t$
        \end{itemize}
        
        \textit{Output.} 
        \begin{itemize}
            \item IPFS Content Identifier: $CID_P$
            \item Cryptographic hash: $H_P = \text{SHA-256}(P)$
            \item On-chain record in Provenance Contract: $\{H_P, CID_P, t\}$
        \end{itemize}
        
        \vspace{0.1cm}
        \hrule
        \vspace{0.1cm}
        
        \textbf{Protocol Phases and Interaction Flow:}
        
        \begin{center}
            \scalebox{0.75}{
            \begin{tikzpicture}[
                actor/.style={rectangle, draw, text width=2.5cm, align=center, minimum height=1.5cm},
                message/.style={->, thick},
                process/.style={rectangle, draw, fill=blue!10, text width=3cm, align=center, rounded corners},
                node distance=1.2cm
            ]
                \node (user) [actor] {Data Owner};
                \node (ipfs) [actor, right=1cm of user] {IPFS Network};
                \node (contract) [actor, right=1cm of ipfs] {Provenance Contract};
                
                \draw[thick, dashed] (user) -- ++(0,-5cm);
                \draw[thick, dashed] (ipfs) -- ++(0,-5cm);
                \draw[thick, dashed] (contract) -- ++(0,-5cm);
                
                \node (step1) [process, below=1cm of user] {Prepare payload $P$};
                \draw[message] (user) -- (step1) node[midway, left] {1};
                
                \node (step2) [process, below=1.2cm of step1] {Compute hash: \\ $H_P = \text{SHA-256}(P)$};
                \draw[message] (step1) -- (step2) node[midway, left] {2};
                
                \node (step3) [process] at (ipfs |- step1) {Store $P$, return: \\ $CID_P$};
                \draw[message, bend left=20] (step1.east) -- node[midway, above] {3} (step3.west);
                
                \node (step4) [process] at (contract |- step2) {Record on-chain: \\ $\{H_P, CID_P, t\}$};
                \draw[message, bend left=20] (step3.east) -| node[pos=0.3, above] {4} (step4.north);
                \draw[message, bend left=20] (step2.east) -- node[pos = 0.7, above] {4} (step4.west);
                
                \node (step5) [process] at ([xshift=1.8cm, yshift=-1cm] user |- step4) {Third-party verification};
                \draw[message, bend left=25] (step4.south) |- node[pos=0.7, below] {5} (step5.east);
            \end{tikzpicture}}
        \end{center} 
        
        \vspace{0.2cm}
        \textbf{Detailed Steps:}
        
        \begin{enumerate}[label=\textbf{\arabic*}, leftmargin=2.5ex, itemsep=0.5ex]
            \item \textbf{Payload Preparation}
                \begin{itemize}
                    \item The data owner prepares the payload $P$ for long-term storage.
                    \item This may involve formatting, encryption, or other preprocessing.
                \end{itemize}
            
            \item \textbf{Hash Computation}
                \begin{itemize}
                    \item Compute the cryptographic hash of the payload using SHA-256:
                    \[
                    H_P = \text{SHA-256}(P).
                    \]
                    \item This hash will be used to verify integrity in the future.
                \end{itemize}
            
            \item \textbf{Off-Chain Storage}
                \begin{itemize}
                    \item The payload $P$ is uploaded to the IPFS network.
                    \item IPFS generates a content identifier $CID_P$ based on the payload's content:
                    \[
                    CID_P = \text{IPFS.Store}(P).
                    \]
                \end{itemize}
            
            \item \textbf{On-Chain Anchoring}
                \begin{itemize}
                    \item The data owner submits the tuple $(H_P, CID_P, t)$ to the Provenance Smart Contract.
                    \item The contract records these values on the blockchain:
                    \[
                    \text{On-chain Record} = \{H_P, CID_P, t, \text{status: "anchored"}\}.
                    \]
                \end{itemize}
            
            \item \textbf{Verification Mechanism}
                \begin{itemize}
                    \item To verify the integrity of payload $P$ at time $t'$:
                    \[
                    \text{Verify}(P, t') \equiv (\text{SHA-256}(P) = H_P) \land (\text{Contract.exists}(H_P, t')).
                    \]
                    \item This ensures both data integrity and immutability.
                \end{itemize}
        \end{enumerate}
    \end{mdframed}
    \caption{The Data Anchoring Protocol}
    \label{prot:data-anchoring}
\end{Protocol}

\subsubsection{Data Anchoring}
The Data Anchoring Protocol (Protocol \ref{prot:data-anchoring}) extends the decentralized provenance framework established by the Task Initiation and State Transition Validation Protocols, providing the infrastructure to ensure long-term integrity for large data payloads (e.g., contracts, audit logs). In particular, the protocol addresses scalability by storing payloads off-chain in IPFS while anchoring their integrity on the blockchain.
The workflow begins with the data owner hashing the payload (\(P\)) using SHA-256 to generate \(H_P\), mirroring the hash computation in task initiation. The payload is then stored on IPFS, yielding a content identifier (\(\text{CID}_P\)) that uniquely maps to \(P\)'s content. Both \(H_P\), \(\text{CID}_P\), and a timestamp (\(t\)) are recorded on the Provenance Smart Contract—reusing the same contract infrastructure as prior protocols—to create an immutable link between the off-chain data and its on-chain proof. This allows third parties to verify \(P\)'s unaltered state at any time by recomputing \(H_P\) and checking its existence in the contract, aligning with the multi-signature verification logic in Protocol \ref{prot:state-validation}.
By separating storage (IPFS) from consensus (blockchain), the protocol efficiently manages large data while leveraging cryptographic guarantees. This design integrates seamlessly with task lifecycles: for example, a milestone in Protocol \ref{prot:state-validation} might reference an anchored dataset, ensuring all stakeholders act on verifiably consistent information.

\begin{Protocol}[]
    \begin{mdframed}[style=ProtocolFrame, font=\small]
        \centering
        \textbf{Authenticated Data Import \& Anchoring Protocol}
        
        \vspace{0.1cm}
        \hrule
        \vspace{0.1cm}
        
        \textit{Participants:} 
        \begin{itemize}[noitemsep, topsep=0pt]
            \item \(\text{A}\): Data Requester (initiates import)
            \item \(\text{O}\): Oracle (verifies provenance)
            \item \(\text{S}\): Data Source (authentic provider)
            \item \(\text{B}\): Blockchain (immutable ledger)
        \end{itemize}
        
        \textit{Core Goals:} 
        \begin{itemize}[noitemsep, topsep=0pt]
            \item Authenticate data \(D\) from \(\text{S}\)
            \item Prove origin/integrity via cryptography
            \item Create auditable blockchain records
            \item Enable decentralized verification (no raw data exposure)
        \end{itemize}
        
        \vspace{0.1cm}
        \hrule
        \vspace{0.1cm}
        
        \textbf{Protocol Phases and Interaction Flow:}
        \begin{center}
            \scalebox{0.9}{
            

            \begin{tikzpicture}[
                entity/.style={circle, draw, minimum size=6mm, font=\scriptsize},
                stepTitle/.style={font=\small\bfseries, anchor=east, thick},
                arrow/.style={->, >=Stealth, semithick, font=\scriptsize}
            ]
            
            \node[stepTitle, align=center] (step1) at (0,0) {1. Mutual Auth \\ \& Key Exchange};
            \node[entity, right=0.5cm of step1] (O) {O};
            \node[entity, right=2cm of O] (A) {A};
            \node[entity, right=2cm of A] (S) {S};
            
            \draw[arrow] (O) -- node[above] {$Y_O = s_O \cdot G$} (A);
            \draw[arrow] (A) -- node[above] {$Y_A = s_A \cdot G + Y_O$} (S);
            
            \node[stepTitle, below=1cm of step1, align=center] (step2) {2. Secure \\ Data Retrieval};
            \node[entity, right=0.5cm of step2] (A2) {A};
            \node[entity, right=2cm of A2] (O2) {O};
            \node[entity, right=2cm of O2] (S2) {S};
            
            \draw[arrow, <->] (A2) -- node[above] {2PC} (O2);
            \draw[arrow] (A2.east) -- ++(0.7cm,0) to[bend right=45] node[pos=0.3, above] {$\hat{Q}$} (S2);
            \draw[arrow] (S2.west) to[bend right=45] node[pos=0.3, below] {$\hat{R}$} ([xshift = -0.7cm] O2.west);
            
            \node[stepTitle, below=1cm of step2, align=center] (step3) {3. Blockchain \\ Anchoring};
            \node[entity, right=0.5cm of step3] (A3) {A};
            \node[entity, right=2cm of A3] (O3) {O};
            \node[entity, right=2cm of O3] (B) {B};
            
            \draw[arrow] (A3) -- node[above] {$\pi$} (O3);
            \draw[arrow] (O3) -- node[above] {$\sigma, h, t$} (B);
            \node[draw, align=left, font=\tiny, below=0.2 of B] {
                $\{h, \text{ID}_S, t, \pi, \sigma, \text{ID}_A\}$
            };
            
            \end{tikzpicture}
                
                }
            \end{center}

        \vspace{0.2cm}
        \textbf{Detailed Steps:}
        
        \begin{enumerate}[label=\textbf{\arabic*.}, leftmargin=2.5ex, itemsep=0.7ex, font=\small]
            \item \textbf{Mutual Authentication \& Key Exchange}
                \begin{itemize}[noitemsep, topsep=0pt]
                    \item \(\text{A}\) initiates TLS with \(\text{S}\), forwards \(\text{S}\)'s certificate to \(\text{O}\) for validation.
                    \item Joint ECDHE key setup:
                        \[
                        \text{O} \xrightarrow{Y_O = s_O \cdot G} \text{A} \xrightarrow{Y_A = s_A \cdot G + Y_O} \text{S}.
                        \]
                    \item Shared secret for collaboration: \(Z = s_S \cdot Y_A = Z_A + Z_O\) (via private keys \(s_A, s_O, s_S\)).
                \end{itemize}
            
            \item \textbf{Secure Data Retrieval}
                \begin{itemize}[noitemsep, topsep=0pt]
                    \item \(\text{A}\) and \(\text{O}\) use 2-party computation (2PC) to construct encrypted query:
                        \[
                        \hat{Q} = \text{AES-CBC}(k_{\text{Enc}}, Q \| \text{HMAC}_{k_{\text{MAC}}}(Q)).
                        \]
                    \item \(\text{S}\) responds with encrypted data \(\hat{R}\); \(\text{A}\) and \(\text{O}\) jointly decrypt and verify integrity via HMAC.
                \end{itemize}
            
            \item \textbf{Blockchain Anchoring with Zero-Knowledge Proof}
                \begin{itemize}[noitemsep, topsep=0pt]
                    \item \(\text{A}\) generates ZKP (\(\pi\)) proving:
                        \[
                        \pi \vdash \{ R \text{ from } \text{S},\, \text{Contextual validity (e.g., } R > \$100) \}.
                        \]
                    \item \(\text{O}\) signs hash \(h = \text{SHA-256}(R)\) and ZKP: \(\sigma = \text{sign}_O(h, \pi, t)\).
                    \item \(\text{B}\) records immutable entry: \(\{h, \text{ID}_S, t, \pi, \sigma, \text{ID}_A\}\).
                    \item Decentralized verification checks: \(h \equiv \text{SHA-256}(R)\), valid \(\pi\), and valid \(\sigma\).
                \end{itemize}
        \end{enumerate}
    \end{mdframed}
    \caption{The Authenticated Data Anchoring Protocol}
    \label{prot:data-import}
\end{Protocol}

\subsubsection{Authenticated Data Import \& Anchoring}
The Authenticated Data Import \& Anchoring Protocol (Protocol \ref{prot:data-import}) extends the cryptographic foundation of the Data Anchoring Protocol (Protocol \ref{prot:data-anchoring}) by introducing authentication and secure data transfer mechanisms for decentralized data ingestion. While the Data Anchoring Protocol focuses on long-term integrity via off-chain storage (IPFS) and on-chain hashing, this protocol addresses the provenance validation phase during data import, ensuring that only authenticated, untainted data is anchored. It achieves this through a three-phase workflow involving mutual authentication, secure data retrieval via 2-party computation (2PC), and blockchain anchoring with zero-knowledge proofs (ZKPs), thereby closing the loop on a tamper-proof data lifecycle from source to storage.

Building on the hash-based anchoring principle of its predecessor, the protocol introduces an Oracle (\(\text{O}\)) as a trust intermediary to validate the data source (\(\text{S}\)) via ECDHE key exchange, ensuring that \(\text{A}\) (data requester) establishes a secure channel with legitimate \(\text{S}\) before retrieval. In particular, Phase 1 establishes mutual authentication and forward-secure keys via ECDHE, ensuring S's legitimacy while preventing key compromise. Phase 2 leverages 2-party computation (2PC) between A and O to securely retrieve encrypted data from S, with HMAC-based integrity checks ensuring tamper-proof transmission. Crucially, Phase 3 integrates with the Data Anchoring Protocol: A generates a zero-knowledge proof (ZKP) attesting to data validity (e.g., contextual thresholds) without exposing raw data, while O cryptographically binds the data hash \(h = \text{SHA-256}(R)\) to S's identity and timestamp \(t\). The final blockchain entry \(\{h, \text{ID}_S, t, \pi, \sigma, \text{ID}_A\}\) integrates the hash \(h\) (analogous to \(H_P\) in the Data Anchoring Protocol) with source metadata (\(\text{ID}_S\)) and cryptographic proofs (\(\pi, \sigma\)), creating a auditable trail that combines the earlier protocol's integrity guarantees with authentication of provenance. This synergy enables a complete ecosystem where data is not only anchored for long-term integrity but also verified at ingestion for authenticity, ensuring end-to-end trust in decentralized data workflows.

\subsection{Smart Contract Layer: Programmable Enforcement of Interaction Rules}
\label{subsec:smart-contract-layer}

The Smart Contract Layer codifies and automates the governance of agent interactions through three key contract archetypes: access control, interaction logic, and identity registration. These contracts collectively enforce protocol compliance, mediate workflows, and dynamically adapt to evolving system requirements, forming the executable backbone of \sys's trust architecture.

\begin{Protocol}[]
    \begin{mdframed}[style=ProtocolFrame, font=\small]
        \centering
        \textbf{Access Control Contract (ACC) Definition}
        
        \vspace{0.1cm}
        \hrule
        \vspace{0.1cm}
        
        Access Control Contracts (ACCs) implement decentralized, context-aware authorization policies to regulate agent privileges. An ACC is formally defined as a tuple:
        \[
        \text{ACC} := \langle \text{resource}, \text{action}, \text{policy} \rangle
        \]
        
        \textbf{Components:}
        
        \begin{enumerate}[label=\textbf{\arabic*}, leftmargin=2.5ex, itemsep=0.5ex]
            \item \textbf{Resource}
                \begin{itemize}
                    \item Specifies a protected asset or data object;
                    \item Examples:
                        \begin{itemize}
                            \item Agent B’s task queue;
                            \item Data repository;
                            \item Smart contract function;
                        \end{itemize}
                    \item Represented as a URI or DID reference.
                    \item 
                \end{itemize}
            
            \item \textbf{Action}
                \begin{itemize}
                    \item Denotes an operation to be performed on the resource;
                    \item Common actions include:
                        \begin{itemize}
                            \item \texttt{read}: Retrieve resource data;
                            \item \texttt{modify}: Update resource state;
                            \item \texttt{invoke}: Execute a smart contract function;
                            \item \texttt{delete}: Remove the resource.
                        \end{itemize}
                \end{itemize}
            
            \item \textbf{Policy}
                \begin{itemize}
                    \item Formalizes authorization rules as a predicate:
                    \[
                    \text{policy}(DID_A, \text{context}) \rightarrow \{\text{true}, \text{false}\}.
                    \]
                    \item Contextual parameters include:
                        \begin{enumerate}[label=\roman*.]
                            \item \textbf{Temporal Constraints}
                                \begin{itemize}
                                    \item Example: \texttt{valid\_after < now < valid\_before};
                                    \item Enables time-bound access permissions;
                                \end{itemize}
                            
                            \item \textbf{DID Attributes}
                                \begin{itemize}
                                    \item Example: \texttt{DID\_A.capabilities.includes("auditor")};
                                    \item Leverages verifiable credentials in the DID document;
                                \end{itemize}
                            
                            \item \textbf{Environmental Variables}
                                \begin{itemize}
                                    \item Example: \texttt{threat\_level < medium};
                                    \item Dynamic factors like network conditions or threat intelligence.
                                \end{itemize}
                        \end{enumerate}
                \end{itemize}
        \end{enumerate}
        
        \vspace{0.2cm}
        \textbf{Policy Evaluation Example:}
        
        \begin{center}
            \scalebox{0.9}{
            \begin{tikzpicture}[
                process/.style={rectangle, draw, fill=blue!10, text width=6cm, align=center, rounded corners},
                node distance=1.2cm
            ]
                \node (policy) [process] {
                    $\text{policy}(DID_A, \text{context}) = $ \\
                    $(\text{context.time} \in [9\,\text{am}, 5\,\text{pm}]) \land$ \\
                    $(\text{DID}_A\,\text{.attributes.role} = \text{"engineer"}) \land$ \\
                    $(\text{context.threatLevel} \leq \text{"medium"})$
                };
            \end{tikzpicture}}
        \end{center}
    \end{mdframed}
    \caption{Access Control Contract (ACC) Formal Definition}
    \label{fig:acc-definition}
\end{Protocol}

\begin{Protocol}[]
    \begin{mdframed}[style=ProtocolFrame, font=\small]
        \centering
        \textbf{Access Control Contract (ACC) Protocol}
        
        \vspace{0.1cm}
        \hrule
        \vspace{0.1cm}
        
        \textit{Input.} 
        \begin{itemize}
            \item Agent $A$'s DID: $DID_A$
            \item Requested action: $\text{action}$
            \item Target resource: $\text{resource}$
            \item Contextual parameters: $\text{context}$
        \end{itemize}
        
        \textit{Output.} 
        \begin{itemize}
            \item Authorization decision: $\{\text{true}, \text{false}\}$
            \item Time-bound capability token (if authorized): 
            \[
            \text{Token}_A = \langle DID_A, \text{action}, \text{resource}, \text{expiry} \rangle
            \]
        \end{itemize}
        
        \vspace{0.1cm}
        \hrule
        \vspace{0.1cm}
        
        \textbf{Protocol Phases and Interaction Flow:}
        
        \begin{center}
            \scalebox{0.75}{
            \begin{tikzpicture}[
                actor/.style={rectangle, draw, text width=3.5cm, align=center, minimum height=1cm},
                message/.style={->, thick},
                process/.style={rectangle, draw, fill=blue!10, text width=3.5cm, align=center, rounded corners},
                node distance=1.2cm
            ]
                \node (agentA) [actor] {Agent $A$};
                \node (acc) [actor, right=1cm of agentA] {Access Control Contract (ACC)};
                \node (resource) [actor, right=1cm of acc] {Protected Resource};
                
                \draw[thick, dashed] (agentA) -- ++(0,-6.5cm);
                \draw[thick, dashed] (acc) -- ++(0,-6.5cm);
                \draw[thick, dashed] (resource) -- ++(0,-6.5cm);
                
                \node (step1) [process, below=0.5cm of agentA] {Request: \\ $\langle DID_A, \text{action}, \text{resource} \rangle$};
                \draw[message] (agentA) -- (step1) node[midway, left] {1};
                
                \node (step2) [process, below=0.5cm of acc] {Retrieve policy for \\ $\langle \text{resource}, \text{action} \rangle$};
                \draw[message, bend left=20] (step1.east) -- node[midway, above] {2} (step2.west);
                
                \node (step3) [process, below=0.5cm of step2] {Evaluate: \\ $\text{policy}(DID_A, \text{context})$};
                \draw[message] (step2) -- (step3) node[midway, left] {3};
                
                \node (step4) [process, below=0.5cm of step3] {Generate token if authorized: \\ $\text{Token}_A = \langle DID_A,$ $\text{action}, \text{resource}, \text{expiry} \rangle$};
                \draw[message] (step3) -- (step4) node[midway, left] {4};
                \draw[message] (step4.west) -- node[midway, above] {4} (agentA |- step4);
                
                \node (step5) [process] at ([yshift = -1cm] resource |- step4) {Authorize access using \\ $\text{Token}_A$};
                \draw[message, bend left=20] (agentA |- step5.west) -- node[pos=0.2, above] {5} (step5.west);
                
                \draw[message, bend left=25] (step5.south) |- node[pos=0.7, above] {6} ([yshift = -0.7cm] agentA |- step5);
            \end{tikzpicture}}
        \end{center} 
        
        \vspace{0.2cm}
        \textbf{Detailed Steps:}
        
        \begin{enumerate}[label=\textbf{\arabic*}, leftmargin=2.5ex, itemsep=0.5ex]
            \item \textbf{Access Request}
                \begin{itemize}
                    \item Agent $A$ sends an access request to the ACC:
                    \[
                    \langle DID_A, \text{action}, \text{resource} \rangle.
                    \]
                \end{itemize}
            
            \item \textbf{Policy Retrieval}
                \begin{itemize}
                    \item The ACC retrieves the corresponding policy for the requested resource and action:
                    \[
                    \text{policy} = \text{ACC.getPolicy}(\text{resource}, \text{action}).
                    \]
                \end{itemize}
            
            \item \textbf{Policy Evaluation}
                \begin{itemize}
                    \item The ACC evaluates the policy against the agent's DID and context:
                    \[
                    \text{authorize}(A, \text{action}) = \text{policy}(DID_A, \text{context}).
                    \]
                    \item Evaluation considers temporal constraints, DID attributes, and environmental factors.
                \end{itemize}
            
            \item \textbf{Token Generation}
                \begin{itemize}
                    \item If the policy evaluation returns \texttt{true}, the ACC generates a capability token: $\text{Token}_A = \langle DID_A, \text{action}, \text{resource}, \text{expiry} \rangle$.
                    
                    \item The token is cryptographically signed by the ACC to prevent forgery.
                \end{itemize}
            
            \item \textbf{Resource Access}
                \begin{itemize}
                    \item Agent $A$ presents the token to the protected resource.
                    \item The resource verifies the token's signature and validity.
                \end{itemize}
            
            \item \textbf{Access Confirmation}
                \begin{itemize}
                    \item The resource returns an access confirmation to Agent $A$.
                    \item The confirmation includes the result of the access attempt.
                \end{itemize}
        \end{enumerate}
    \end{mdframed}
    \caption{The Access Control Contract (ACC) Protocol}
    \label{prot:access-control}
\end{Protocol}

\subsubsection{Access Control Contract}

The \textbf{Access Control Contract (ACC)} represents a decentralized authorization framework designed for multi-agent interations in distributed environments. As shown in Figures~\ref{fig:acc-definition} and~\ref{prot:access-control}, the ACC combines verifiable credentials with dynamic context evaluation to enable fine-grained, adaptive access control.

\parab{Contract Components}  
The ACC is structured as a triple $\langle \text{resource},$ $\text{action}, \text{policy} \rangle$ that establishes verifiable relationships between decentralized identifiers (DIDs) and protected assets. The resource specification protects specific digital assets through URI/DID references, including agent-specific resources (e.g., task queues and data stores), smart contract interfaces, and cryptographic key material. Action typing prevents privilege escalation through strict operation typing, with core actions constrained at the protocol level to \texttt{read}, \texttt{modify}, \texttt{invoke}, and \texttt{delete}. Contextual policies incorporate three dimensions of authorization predicates: temporal constraints using logical clock comparisons, DID metadata verification of agent credentials and capabilities, and environmental signals from real-time inputs such as threat intelligence feeds or network load metrics.

\parab{Authorization Protocol}  
The ACC protocol implements a stateful capability system through six-phase interaction flows. First, agents compose requests as $\langle DID, \text{action}, \text{resource} \rangle$ tuples with current context metadata. The ACC then resolves governing policies through content-addressable storage using $\text{(resource, action)}$ composite keys. Policy execution evaluates both static DID attributes and dynamic variables through first-order logic predicates:

\[
\phi(DID, t, \theta) = \bigwedge_{i} \left[ \psi_i(DID) \land \gamma_i(t) \land \xi_i(\theta) \right]
\]

where $\psi$, $\gamma$, and $\xi$ represent DID, temporal, and environmental sub-predicates respectively. Approved requests generate capability tokens containing action-resource bindings with expiration times, cryptographic authorization proofs, and context snapshots for replay protection. Protected resources subsequently validate tokens through signature checks and context recency verification. Finally, successful accesses update resource-specific usage metrics that feed back into policy evaluations through state synchronization.

\parab{Security Properties}  
The ACC framework guarantees three fundamental security properties: \first \textit{least privilege} through capability tokens granting only specific access rights, \second \textit{temporal restriction} via mandatory expiration timestamps on all tokens, and \third \textit{non-repudiation} achieved by cryptographically binding tokens to both issuing ACC and requesting DID. This architecture enables dynamic privilege management in decentralized systems while maintaining auditability through on-chain policy execution records. The context-aware design supports adaptive security postures that respond to real-time operational conditions.

\parab{Advantages Over Traditional Access Control Models}  
The ACC framework provides three key improvements over conventional systems. First, its dynamic policy enforcement enables real-time adaptation to contextual factors like temporal constraints and threat levels, overcoming the rigidity of static role-based access control (RBAC) lists. Second, ACC achieves fine-grained permissions through atomic $\langle$resource, action$\rangle$ bindings, eliminating broad privilege grants common in attribute-based access control (ABAC). Finally, the decentralized policy management using standardized DIDs/URIs bridges semantic gaps across organizational boundaries, contrasting with centralized policy servers that create institutional silos. These features combine to support secure collaboration in decentralized ecosystems while maintaining auditability – a critical requirement absent in many traditional models relying on implicit trust boundaries.

\begin{Protocol}[]
    \begin{mdframed}[style=ProtocolFrame, font=\small]
        \centering
        \textbf{Interaction Logic Contract (ILC) Definition}
        
        \vspace{0.1cm}
        \hrule
        \vspace{0.1cm}
        
        Interaction Logic Contracts (ILCs) encode domain-specific workflows as deterministic state machines, ensuring protocol adherence. An ILC is formally defined as a tuple:
        \[
        \text{ILC} := \langle \Sigma, s_0, \delta, G \rangle
        \]
        
        \textbf{Components:}
        
        \begin{enumerate}[label=\textbf{\arabic*}, leftmargin=2.5ex, itemsep=0.5ex]
            \item \textbf{State Set ($\Sigma$)}
                \begin{itemize}
                    \item Finite set of workflow states
                    \item Examples:
                        \begin{itemize}
                            \item \texttt{OrderCreated}, \texttt{ProductionConfirmed}, \texttt{Shipped}
                            \item \texttt{ProposalSubmitted}, \texttt{VotingActive}, \texttt{DecisionReached}
                        \end{itemize}
                    \item Represented as human-readable identifiers
                \end{itemize}
            
            \item \textbf{Initial State ($s_0$)}
                \begin{itemize}
                    \item Starting state of the state machine
                    \item Example: \texttt{OrderCreated} for a supply chain workflow
                \end{itemize}
            
            \item \textbf{State Transition Function ($\delta$)}
                \begin{itemize}
                    \item Defines state changes triggered by events:
                    \[
                    \delta: \Sigma \times E \rightarrow \Sigma
                    \]
                    \item $E$ is the set of valid events (e.g., \texttt{Confirm}, \texttt{Approve}, \texttt{Reject})
                    \item Example transition: 
                    \[
                    \delta(\texttt{OrderCreated}, \texttt{PaymentReceived}) = \texttt{ProductionScheduled}
                    \]
                \end{itemize}
            
            \item \textbf{Transition Guards ($G$)}
                \begin{itemize}
                    \item Preconditions for state transitions:
                    \[
                    G: \Sigma \rightarrow \text{Guard}
                    \]
                    \item Guards can include:
                        \begin{enumerate}[label=\roman*.]
                            \item Multi-signature requirements (e.g., $\geq 2$ approvals from 3 participants)
                            \item Time constraints (e.g., transition must occur within 72 hours)
                            \item Data validity checks (e.g., hash matches on-chain record)
                        \end{enumerate}
                    \item Example guard: 
                    \[
                    G(\texttt{ProductionScheduled}) = (\text{number of approvals} \geq quorum)
                    \]
                \end{itemize}
        \end{enumerate}
        
        \vspace{0.2cm}
        \textbf{State Machine Example:}
        
        \begin{center}
            \scalebox{0.75}{
            \begin{tikzpicture}[
                state/.style={ellipse, draw, minimum height=1cm, font=\small, fill=blue!5},
                transition/.style={->, thick, shorten >=2pt, shorten <=2pt},
                guard/.style={rectangle, draw, fill=blue!10, rounded corners, font=\scriptsize, align=center, text width=3cm},
                initial/.style={-latex, shorten >=2pt}
            ]
                \node (created) [state] {\texttt{OrderCreated}};
                \node (scheduled) [state, right=4cm of created] {\texttt{ProductionScheduled}};
                \node (shipped) [state, below=2cm of scheduled] {\texttt{Shipped}};
                
                \draw[transition] (created) -- node[above, guard] {Event: \texttt{PaymentReceived} \\ Guard: $amount > 0$} (scheduled);
                \draw[transition] (scheduled) -- node[left, guard] {Event: \texttt{ManufacturingComplete} \\ Guard: $QC\_passed = \text{true}$} (shipped);
                
                \draw[initial] (0,-1.5cm) -- (created.south);
                
                \node[above=0.2cm of created] (s1) {\textbf{State 1}};
                \node[above=0.2cm of scheduled] (s2) {\textbf{State 2}};
                \node[left=0.2cm of shipped] (s3) {\textbf{State 3}};
            \end{tikzpicture}}
        \end{center}
        
    \end{mdframed}
    \caption{Interaction Logic Contract (ILC) Formal Definition}
    \label{fig:ilc-definition}
\end{Protocol}

\begin{Protocol}[]
    \begin{mdframed}[style=ProtocolFrame, font=\small]
        \centering
        \textbf{Interaction Logic Contract (ILC) Protocol}
        
        \vspace{0.1cm}
        \hrule
        \vspace{0.1cm}
        
        \textit{Input.} 
        \begin{itemize}
            \item Initial state $s_0 \in \Sigma$; State transition function $\delta: \Sigma \times E \rightarrow \Sigma$
            \item Transition guards $G: \Sigma \rightarrow \text{Guard}$
            \item Event $e \in E$ triggering state transition
        \end{itemize}
        
        \textit{Output.} 
        \begin{itemize}
            \item Updated state $s_j = \delta(s_i, e)$
            \item Execution receipts and on-chain records of state transition
        \end{itemize}
        
        \vspace{0.1cm}
        \hrule
        \vspace{0.1cm}
        
        \textbf{Protocol Phases and Interaction Flow:}
        
        \begin{center}
            \scalebox{0.75}{
            \begin{tikzpicture}[
                actor/.style={rectangle, draw, text width=2.5cm, align=center, minimum height=1cm},
                message/.style={->, thick},
                process/.style={rectangle, draw, fill=blue!10, text width=3cm, align=center, rounded corners},
                node distance=1.2cm
            ]
                \node (agents) [actor] {Participants};
                \node (ilc) [actor, right=1cm of agents] {ILC Smart Contract};
                \node (ledger) [actor, right=1cm of ilc] {Blockchain Ledger};
                
                \draw[thick, dashed] (agents) -- ++(0,-5.5cm);
                \draw[thick, dashed] (ilc) -- ++(0,-5.5cm);
                \draw[thick, dashed] (ledger) -- ++(0,-5.5cm);
                
                \node (init1) [process, below=0.5cm of agents] {Jointly define: \\ $\Sigma, \delta, G, s_0$};
                \draw[message] (agents) -- (init1) node[midway, left] {1};
                
                \node (init2) [process] at (ilc |- init1) {Deploy ILC contract};
                \draw[message] (init1.east) -- node[midway, above] {1} (init2.west);
                
                \node (step1) [process, below=0.5cm of init1] {Collect signatures for event $e$};
                \draw[message] (init1) -- (step1) node[midway, left] {2};
                
                \node (step2) [process] at (ilc |- step1) {Verify: \\ $G(s_i) \land \text{signatures} \geq k$};
                \draw[message, bend left=20] (step1.east) -- node[midway, above] {3} (step2.west);
                
                \node (step3) [process, below=0.5cm of step2] {Compute: \\ $s_j = \delta(s_i, e)$};
                \draw[message] (step2) -- (step3) node[midway, left] {4};
                
                \node (step4) [process, below=0.5cm of ledger) [process, below=1cm of ledger] {Record: \\ $(s_i, e, s_j, t)$};
                \draw[message, bend left=20] (step3.east) -- ++(0.2cm, 0) |- node[pos=0.3, right] {5} (step4.west);
                
                \node (step5) [process, below=0.5cm of step3] {Return execution receipt};
                \draw[message] (step4.south) |- node[pos=0.3, right] {6} (step5.east);
                
                \draw[message, bend left=25] (step5.west) -- node[midway, below] {6} (agents |- step5);
            \end{tikzpicture}}
        \end{center} 
        
        \vspace{0.2cm}
        \textbf{Detailed Steps:}
        
        \begin{enumerate}[label=\textbf{\arabic*}, leftmargin=2.5ex, itemsep=0.5ex]
            \item \textbf{ILC Initialization}
                \begin{itemize}
                    \item Participants collaboratively define:
                        \begin{itemize}
                            \item State set $\Sigma$; Transition function $\delta$;
                            \item Guards $G$; Initial state $s_0$.
                        \end{itemize}
                    \item Deploy the ILC smart contract encoding these parameters.
                \end{itemize}
            
            \item \textbf{Event Submission}
                \begin{itemize}
                    \item For a transition $s_i \xrightarrow{e} s_j$, participants:
                        \begin{enumerate}[label=\alph*.]
                            \item Collect $k$ signatures for event $e$, e.g., production confirmation signed by 3/5 authorized agents.
                        \end{enumerate}
                \end{itemize}
            
            \item \textbf{Guard Verification}
                \begin{itemize}
                    \item The ILC contract verifies:
                        \begin{enumerate}[label=\alph*.]
                            \item Multi-signature threshold ($\geq k$ valid signatures);
                            \item Consistency with on-chain task hashes:
                                \[
                                H_{\text{task}} = \text{SHA-256}(\text{metadata} \| \text{participants} \| t);
                                \]
                            \item Any additional contextual guards (e.g., time constraints).
                        \end{enumerate}
                \end{itemize}
            
            \item \textbf{State Transition Execution}
                \begin{itemize}
                    \item If guards pass, the ILC computes the new state:
                        \[
                        s_j = \delta(s_i, e).
                        \]
                    \item Emits an event logging the transition:
                        \[
                        \text{TransitionEvent}(s_i, e, s_j, \text{timestamp}).
                        \]
                \end{itemize}
            
            \item \textbf{Ledger Recording}
                \begin{itemize}
                    \item The blockchain records the following information to ensure immutability and auditability:
                        \begin{itemize}
                            \item Previous state $s_i$; Triggering event $e$;
                            \item New state $s_j$; Transaction timestamp $t$.
                        \end{itemize}
                \end{itemize}
            
            \item \textbf{Receipt Distribution}
                \begin{itemize}
                    \item The ILC sends execution receipts (including transaction hash and new state) to participants.
                \end{itemize}
        \end{enumerate}
    \end{mdframed}
    \caption{The Interaction Logic Contract (ILC) Protocol}
    \label{prot:ilc-execution}
\end{Protocol}

\subsubsection{Interaction Logic Contract}

The \textbf{Interaction Logic Contract (ILC)} formalizes multi-party workflows as blockchain-enforced state machines, providing unified task views and deterministic execution of complex business logic across decentralized networks. As detailed in Protocols~\ref{fig:ilc-definition} and~\ref{prot:ilc-execution}, ILCs combine finite state machines with cryptographic verification to ensure protocol compliance in cross-organizational interactions.

\parab{Contract Components}
The ILC is mathematically defined as $\langle \Sigma, s_0, \delta, G \rangle$ where $\Sigma$ represents the state set encoding domain-specific workflow milestones through human-readable labels such as \texttt{ProductionConfirmed} and \texttt{VotingActive}, serving dual purposes as coordination points and audit checkpoints. The transition function $\delta$ enforces deterministic state changes through event-triggered transitions governed by the condition:

\[
\delta(s_i, e) = s_j \iff \exists \text{valid path } s_i \xrightarrow{e} s_j \text{ in workflow DAG}
\]

preventing invalid state jumps through algorithmic enforcement. Transition preconditions $G$ implement context-aware validation through three complementary mechanisms: consensus requirements expressed as $k$-of-$n$ signature thresholds for collective decisions, temporal constraints defining valid time windows for transitions, and data integrity checks using cryptographic hashes to verify off-chain task completion.

\parab{Workflow Execution Protocol}  
The ILC operationalizes workflow management through a six-phase lifecycle. Cooperative initialization begins with participants jointly defining state machine parameters via multi-signature deployment, establishing shared protocol semantics. During event orchestration, authorized actors collect digital signatures for state transition requests, with signature thresholds enforcing governance policies. Guard verification occurs through smart contract validation of:

\[
G(s_i) \triangleq \left(\text{sigCount} \geq k\right) \land \left(t \in [t_{start}, t_{end}]\right) \land \left(H_{\text{task}} == H_{\text{on-chain}}\right)
\]

State advancement follows through deterministic transition execution updating workflow status via $s_j = \delta(s_i, e)$ where $e \in E$ represents signed event data. The blockchain ledger then immutably records transition tuples $\langle s_i, e, s_j, t\rangle$, creating non-repudiable audit trails. Finally, cryptographic receipts distribute state change confirmations and transaction proofs to all stakeholders through participant notification.

\parab{Advantages Over Traditional Workflow Systems}  
ILCs address three critical limitations of conventional systems by replacing centralized orchestration engines with decentralized state transition logic, eliminating single points of control. They enable provable compliance with business rules through cryptographic guards, contrasting with opaque policy enforcement in Business Process Management (BPM) systems. The blockchain-anchored state machine creates tamper-evident process histories, solving audit challenges in multi-jurisdictional operations. These properties make ILCs particularly suitable for supply chain coordination, decentralized governance, and regulated multi-party processes requiring strong compliance guarantees.

\parab{Security Guarantees}  
The ILC architecture provides three foundational security properties: \first \textit{state integrity} through cryptographic hash chaining, preventing retrospective modifications, \second \textit{transition finality} ensuring state changes achieve Byzantine fault-tolerant consensus once recorded on-chain, and \third \textit{non-repudiation} via digital signatures binding participants to specific transitions. This combination of formal state modeling and blockchain enforcement enables trustless collaboration between mutually distrusting parties while maintaining operational flexibility through configurable guard conditions. The ILC's deterministic execution model bridges the gap between rigid smart contract logic and real-world business process variability.

\begin{Protocol}[]
    \begin{mdframed}[style=ProtocolFrame, font=\small]
        \centering
        \textbf{Agent Governance Contract (AGC) Definition}
        
        \vspace{0.1cm}
        \hrule
        \vspace{0.1cm}
        
        Agent Governance Contracts (AGCs) manage the lifecycle of DIDs and capabilities, serving as the root of trust for identity operations. An AGC implements four core functions:
        
        \begin{enumerate}[label=\textbf{\arabic*}, leftmargin=2.5ex, itemsep=0.5ex]
            \item \textbf{Registration}
                \begin{itemize}
                    \item Maps a new DID to its initial capabilities:
                    \[
                    \text{register}(DID_A, H(D_A)) \rightarrow \text{emits DIDCreated}(DID_A);
                    \]
                    \item Initializes the DID with a hash of its initial document $D_A$.
                \end{itemize}
            
            \item \textbf{Capability Update}
                \begin{itemize}
                    \item Authorizes modifications to $D_A$.capabilities via multi-signature approval:
                    \[
                    \text{update}(DID_A, H(D_A')) \quad \text{s.t.} \quad \text{validUpdatePolicy}(\Delta C);
                    \]
                    \item $\Delta C$ represents the capability delta (e.g., adding image\_recognition);
                    \item Requires sufficient signatures from authorized controllers.
                \end{itemize}
            
            \item \textbf{Revocation}
                \begin{itemize}
                    \item Invalidates a DID upon security incidents:
                    \[
                    \text{revoke}(DID_A) \quad \text{if} \quad \text{isAuthorized}(\text{Revoker}, DID_A);
                    \]
                    \item Ensures only authorized entities can revoke the DID.
                \end{itemize}
            
            \item \textbf{Lookup}
                \begin{itemize}
                    \item Resolves DIDs to current metadata hashes:
                    \[
                    \text{resolve}(DID_A) \rightarrow H(D_A);
                    \]
                    \item Returns the latest hash of the DID document.
                \end{itemize}
        \end{enumerate}
        
        \vspace{0.2cm}
        \textbf{State Machine Example:}
        
        \begin{center}
            \scalebox{0.75}{
            \begin{tikzpicture}[
                state/.style={ellipse, draw, minimum height=1cm, font=\small, fill=blue!5},
                transition/.style={->, thick, shorten >=2pt, shorten <=2pt},
                guard/.style={rectangle, draw, fill=blue!10, rounded corners, font=\scriptsize, align=center, text width=3.5cm},
                initial/.style={-latex, shorten >=2pt}
            ]
                \node (created) [state] {\texttt{Created}};
                \node (active) [state, right=4.5cm of created] {\texttt{Active}};
                \node (revoked) [state, below=2cm of active] {\texttt{Revoked}};
                
                \draw[transition] (created) -- node[above, guard] {Event: \texttt{Register} \\ Guard: $validDIDFormat(DID_A)$} (active);
                \draw[initial] (active.east) ++(5cm,0) -- node[above, guard] {Event: \texttt{Update} \\ Guard: $quorum\_met$} (active.east);
                \draw[transition] (active) -- node[right, guard] {Event: \texttt{Revoke} \\ Guard: $isAuthorized$} (revoked);
                
                \draw[initial] (0,-1.5cm) -- (created.south);
                
                \node[above=0.2cm of created] (s1) {\textbf{State 1}};
                \node[above=0.2cm of active] (s2) {\textbf{State 2}};
                \node[left=0.2cm of revoked] (s3) {\textbf{State 3}};
            \end{tikzpicture}}
        \end{center}
        
    \end{mdframed}
    \caption{Agent Governance Contract (AGC) Formal Definition}
    \label{fig:AGC-definition}
\end{Protocol}

\subsubsection{Agent Governance Contract}

The \textbf{Agent Governance Contract (AGC)} establishes an agent-profiling system for multi-agent interactions, managing the complete lifecycle of agents' decentralized identifiers (DIDs) and associated capabilities. As shown in Protocol~\ref{fig:AGC-definition}, AGCs combine cryptographic identity management with state machine enforcement to create non-repudiable audit trails for agent identity operations.

\parab{Contract Components}
The AGC implements four fundamental operations through constrained state transitions. Provisioning trust anchors initializes DIDs with cryptographic proof of control through the registration operation $\text{register}(DID_A, H(D_A)) \rightarrow \text{rootHash}[DID_A] = H(D_A)$, simultaneously emitting \texttt{DIDCreated} events with initial capability commitments. Dynamic capability management enforces multi-signature update policies defined by $\text{validUpdate} \triangleq \sum_{i=1}^n \text{sig}_i(C') \geq k \land H(D_A') \neq H(D_A)$, maintaining versioned capability sets through hash chaining. Emergency revocation invalidates compromised DIDs via authorized triggers following $\text{revoke}(DID_A) \rightarrow \text{state}[DID_A] = \texttt{Revoked} \iff$ $\text{authZ}(\text{Revoker}, \text{AGC\_ADMIN})$. Decentralized resolution provides tamper-proof verification through the $\text{resolve}(DID_A) \rightarrow \langle H(D_A),$ $\text{state}, \text{validFrom}, \text{validUntil} \rangle$ operation, enabling cryptographic proof of DID document validity.

\parab{Workflow Execution Protocol}  
The AGC lifecycle progresses through five constrained phases. First, during DID creation, agents submit \texttt{register} transactions with initial capability hashes, with the AGC verifying DID syntax per W3C standards before transitioning to \texttt{Active} state. Capability management follows, where authorized controllers propose updates via multi-signature bundles validated against the conditions $\text{quorumMet} \land \text{timeLockActive} \land \neg\text{isRevoked}(DID_A)$. The revocation process enables security operators to trigger permanent state transitions to \texttt{Revoked} using admin credentials, setting non-clearable flags. For DID resolution, clients query current state and metadata through \texttt{resolve} calls that return Merkle proofs of document hashes. Finally, state monitoring involves watchdog agents tracking changes through \texttt{DIDEvent} logs and enforcing SLAs via heartbeat checks on \texttt{Active} DIDs.

\parab{Advantages Over Traditional Identity Systems}  
AGCs address three critical limitations of conventional systems: \first enabling self-sovereign capability management through decentralized multi-signature updates that eliminate centralized certificate authorities, \second providing temporal validity enforcement via state machines where static DNS-based systems fail, and \third establishing cryptographic non-repudiation for all identity operations through blockchain-anchored hashing - a capability absent in traditional OAuth/JWT frameworks. These advancements particularly benefit decentralized autonomous organizations (DAOs) and IoT networks requiring granular, auditable access control over dynamic participant sets.

\parab{Security Properties}  
The AGC architecture guarantees three foundational security properties: \first \textit{Non-repudiable operations} through cryptographically signed state transitions recorded on-chain, \second \textit{Least-privilege updates} enforced via capability modification diffs with monotonicity checks, and \third \textit{Revocation finality} preventing zombie identity attacks by permanently disabling revoked DIDs. This combination of cryptographic state management and blockchain-enforced lifecycle controls enables secure agent coordination at scale while maintaining operational flexibility through configurable governance rules.

\subsubsection{Integration with Prior Layers}

The Smart Contract Layer interoperates with \sys’s identity and Ledger Layers through mechanisms including:
\begin{itemize}
  \item Cross-Layer Validation: Access control decisions (from ACCs) incorporate real-time DID status checks via AGC lookups and ledger-layer state proofs.
  \item Ability Extension: AGCs and ILCs essentially extend the abilities of the DID Registry Contract and Provenance Contract required in the Identity Layer and the Ledger Layer, by enabling agent lifecycle management and automatic task-flow enforcement. Instead of providing simple immutable records and traceability, AGCs and ILCs step further to fully leverage the foundation of blockchain to rigorously ensure behavior/protocol compliance in cross-organizational agent interactions.  
  \item Event-Driven Syncing: Contract state changes (e.g., capability updates) trigger ledger-layer events (e.g., anchoring new hash), ensuring system-wide consistency.
\end{itemize}

This design ensures all agent interactions are governed by transparent, auditable, and tamper-proof rules—shifting trust from centralized authorities to mathematically verifiable protocols. By unifying access control, workflow logic, and identity lifecycle management, the Smart Contract Layer enables \sys to scale securely across open and dynamic ecosystems.

\section{Instantiation of \sys}
\label{sec:instantiation-framework}

In this section, we present a formal and general framework for instantiating \sys within various existing MASes, including but not limited to the MASes following Supervisor-Based, Network/Graph-Based, and Federated Learning-Based collaboration models. The goal is to not only specify the efforts required to adapt \sys to a specific MAS, but also provide a standard methodology or easy-to-practice guideline for fully leveraging \sys to secure various MASes.  

\subsection{Instantiation Principles}

Our framework of instantiating \sys adheres to the following design principles:

\begin{itemize}
    \item \textit{Rigorous Protocol Translation}: To bridge the semantics of an existing MAS and that of \sys, a set of \textit{transformation functions} should be rigorously defined to map MAS-specific data and protocols to \sys's canonical formats. For example, transformation functions are responsible for migrating MAS-specific identifiers to DIDs, and transforming MAS-specific task metadata to the DID-driven \sys task metadata.

    \item \textit{Modular and Pluggable Integration}: Since the core \sys layers (Identity, Ledger, Smart Contract layer with corresponding protocols) are modularly designed, selective layer instantiation within an existing MAS is feasible. For example, the user MAS of \sys might choose to instantiate only the Identity and the Ledger layer to enable accountability of agent interaction history, without relying on the Smart Contract layer for access control or automating interaction logic.
    
    \item \textit{Trust Preservation}: MASes following our instantiation framework to integrate \sys (same components) shall offer the same level of authenticity, integrity and accountability. This ensures that trust established in one MAS can be validated and leveraged across others, thereby enabling cross-MAS interactions.
\end{itemize}

\subsection{Transformation Functions}

Let $\mathcal{P}$ denote the set of MASes based on various collaboration paradigms. For each MAS $P \in \mathcal{P}$, the instantiation of \sys defines:

\begin{enumerate}
    \item \textit{Identity Mapping Function}: $\mathcal{I}_P: \text{ID}_P \rightarrow \text{DID},$, where $\text{ID}_P$ is the paradigm-specific identity space.
    For example, in a federated learning-based MAS, $\text{ID}_{\text{FL}}$ might consist of client IDs and institutional profiles, which are mapped to DIDs and DID documents through $\mathcal{I}_{\text{FL}}$. $\mathcal{I}_P$ could be constructed based on \sys's Legacy Identity Migration protocol (see Protocol~\ref{prot:legacy-to-did-migration-part2}).
    
    \item \textit{Metadata Transformation}: $\mathcal{T}_P: \text{Metadata}_P \rightarrow \text{Metadata}_{\sys},$ which transforms paradigm-specific task meta information into \sys compatible form.
    For instance, in a graph-based workflow, $\mathcal{T}_{\text{Graph}}$ would transform a DAG node's metadata (e.g., "image processing task") into \sys's task format, ensuring all required fields (e.g., performer DID, input/output specifications) are properly populated and anchored on chain.
    
    \item \textit{Protocol Translation}: $\mathcal{A}_P: \text{Op}_P \times \text{State}_P \rightarrow \text{Transaction}_{\sys},$ where $\text{Op}_P$ are paradigm operations and $\text{State}_P$ is internal state.
    In a supervisor-based system, for example, the operation of aggregating multiple agents' answers ($\text{Op}_{\text{Supervisor}}$) would be translated into a state-transition (see Protocol~\ref{prot:state-validation}) transaction in \sys, which records the aggregation process, the final result and requires a multi-signature for consensus.
\end{enumerate}

\subsection{Layer-Specific Instantiation}

With the help of the transformation functions defined above, here we formally introduce the layer-specific instantiation of \sys.

\parab{Identity Layer Instantiation}
Formally defined as a tuple:
\[
\text{IdentityInst} = \langle \mathcal{I}, \mathcal{S}, \mathcal{V}, \mathcal{R} \rangle,
\]
where:
\begin{itemize}
    \item $\mathcal{I}: \text{ID}_{P \in \mathcal{P}} \rightarrow \text{DID}$ is the identity mapping function;
    \item $\mathcal{S}: \text{PrivateKey} \times \text{Message} \rightarrow \text{Message}_S \times \text{DID}$ generates DID-signed message with the corresponding DID;
    \item $\mathcal{V}: \text{DID} \times \text{Signature} \times \text{Message}_S \rightarrow \{\top, \bot\}$ verifies DID-bound messages;
    \item $\mathcal{R}: \text{DID} \rightarrow \text{PublicKey} \times \text{Attributes}$ resolves DID documents.
\end{itemize}

\textbf{Example}: When instantiating the Identity Layer of \sys within a supervisor-based MAS, an agent identified by a legacy username ($\text{ID}_{\text{Legacy}}$) is mapped to a DID through $\mathcal{I}$. When this agent sends a message to the supervisor, $\mathcal{S}$ signs the message via its DID-bound private key, while $\mathcal{V}$ verifies the message signature using the public key resolved via $\mathcal{R}$.

\parab{Ledger Layer Instantiation}
Defined as:
\[
\text{LedgerInst} = \langle \mathcal{T}_{\text{task}}, \mathcal{A}_{\text{state}}, \mathcal{A}_{\text{anchor}} \rangle,
\]
with component functions:
\begin{itemize}
    \item $\mathcal{T}_{\text{task}}: \text{Metadata}_{P \in \mathcal{P}} \rightarrow \text{Metadata}_{\sys}$ constructs \sys-compatible task metadata, which includes the initiator DID, participant DIDs and deadline, based on the original MAS metadata;
    \item $\mathcal{A}_{\text{state}}: \text{Op}_{P \in \mathcal{P}} \times \text{State}_P \rightarrow \text{Transaction}_{\sys}$ composes the blockchain transaction for anchoring the transition of task states;
    \item $\mathcal{A}_{\text{anchor}}: \text{Data}_{P \in \mathcal{P}} \rightarrow \text{CID} \times \text{Hash} \times \text{Timestamp}$ adds on-chain anchoring and integrity verification of large payloads.
\end{itemize}

\textbf{Example}: In a MAS that instantiates \sys, when a task is completed, $\mathcal{T}_{\text{state}}$ transforms the paradigm-specific task state transition (e.g., "image processed") into a \sys state transition transaction, recorded on chain. Task-specific large payloads are also anchored on chain via $\mathcal{A}_{\text{anchor}}$. This ensures the task's progress and relevant data are verifiably tracked.

\parab{Smart Contract Layer Instantiation}
Formalized as:
\[
\text{ContractInst} = \langle \mathcal{C}_{\text{ACC}}, \mathcal{C}_{\text{ILC}}, \mathcal{C}_{\text{ARC}} \rangle,
\]
where each component generates smart contracts (with corresponding interaction protocols) tailored to MAS-specific requirements:
\begin{itemize}
    \item $\mathcal{C}_{\text{ACC}}: \text{Policy}_P \rightarrow \text{ACC}, Protocol_{ACC};$
    \item $\mathcal{C}_{\text{ILC}}: \text{Workflow}_P \rightarrow \text{ILC}, Protocol_{ILC};$
    \item $\mathcal{C}_{\text{ARC}}: \text{Capabilities}_P \rightarrow \text{ARC}, Protocol_{ARC}.$
\end{itemize}

\textbf{Example}: The MAS-specific access policy (e.g., "only PhD researchers can access model weights") is translated into an ACC via $\mathcal{C}_{\text{ACC}}$, which rigorously define the resources and requestor attributes in the ACC policies. After deploying ACC, the MAS can interact with it through $Protocol_{ACC}$ (specified in Protocol~\ref{prot:access-control}) to enforce resource access control.

\subsection{Trust Preservation Theorem}

\begin{theorem}[Cross-MAS Trust Equivalence]
Given a valid interaction $\mathcal{I}_P$ in the MAS $P \in \mathcal{P}$, the instantiation of \sys within $P$, which transforms $\mathcal{I}_P$ into $\mathcal{A}_P(\mathcal{I}_P)$, preserves the following properties:
\begin{enumerate}
    \item \textbf{Authenticity}: If $\mathcal{I}_P$ is authentic under $P$'s rules, $\mathcal{A}_P(\mathcal{I}_P)$ is verifiable via \sys's Identity Layer.
    \item \textbf{Integrity}: The integrity of data and state transitions in $\mathcal{I}_P$ is preserved in $\mathcal{A}_P(\mathcal{I}_P)$ as verified by the Ledger Layer.
    \item \textbf{Accountability}: Every action in $\mathcal{I}_P$ is traceable to an identified agent in \sys's identity system.
\end{enumerate}
\label{theorem:trust-equivalence}
\end{theorem}

\begin{proof}[Sketch of Proof]
The proof follows from the rigorous design of the transformation functions $\mathcal{I}$, $\mathcal{T}$, and $\mathcal{A}$, which ensure that each paradigm-specific interaction is translated into a verifiable sequence of \sys transactions. DID-based identity verification, on-chain hash anchoring, and smart contract enforcement jointly guarantee the preservation of authenticity, integrity, and accountability.

\textbf{Authenticity}: By construction, $\mathcal{I}_P$ maps all paradigm identities to DIDs, and $\mathcal{V}$ verifies all interactions using these DIDs. Thus, if an interaction is authentic in $P$, it will be verifiable in \sys.

\textbf{Integrity}: $\mathcal{T}_P$ ensures that metadata and state transitions are semantically preserved, and $\mathcal{A}_{\text{anchor}}$ uses cryptographic hashing to anchor data to the ledger, preventing tampering.

\textbf{Accountability}: All actions are recorded in the ledger with associated DIDs, ensuring traceability back to responsible agents.
\end{proof}

\subsection{Security and Performance Analysis}

\parab{Security Guarantees}
Essentially, our framework of instantiating \sys within an existing MAS provides the following \textit{additional} security properties:
\begin{itemize}
    \item \textit{Unified Identity with Message Authentication}: Agents' MAS-specific identities are enhanced with DIDs, enabling cross-domain interoperability and DID-driven communication protection..
    \item \textit{Data Integrity}: All data (either large payloads or interaction history) is cryptographically anchored on the ledger, ensuring tamper-evidence.
    \item \textit{Access Control}: Smart contracts enforce fine-grained and dynamically configurable access policies, preventing unauthorized actions.
    \item \textit{Non-Repudiation}: Cryptographic signatures and on-chain records prevent agents from denying their actions.
\end{itemize}

\parab{Performance Considerations}
While the instantiation of \sys introduces some overhead to existing MASes (evaluated in \S~\ref{sec:evaluation}), promising optimizations include:
\begin{itemize}
    \item \textit{Off-Chain Full Data Storage}: Large datasets are stored off-chain (e.g., IPFS, reliable cloud storage) with only hashes and metadata on-chain.
    \item \textit{Layer-2 Scaling}: For high-throughput paradigms, Layer-2 solutions (e.g., ZK-Rollups) can be used to reduce blockchain congestion.
    \item \textit{Caching Mechanisms}: Frequently accessed data (e.g., DID documents, access control tokens) are cached locally to reduce network latency.
\end{itemize}


\section{Defense Orchestration Engine}
\label{sec:doe}

The Defense Orchestration Engine (DOE) leverages BlockA2A’s three-layer architecture to deliver proactive threat detection, automated response, and forensic analysis capabilities, ensuring robust protection against malicious activities within the ecosystem. By integrating real-time monitoring, advanced analytics, and adaptive response mechanisms, the engine maintains continuous vigilance over system operations, identifies anomalies, and enforces security policies to mitigate risks.

\begin{figure}[ht]
    \centering
    \includegraphics[width=0.8\columnwidth]{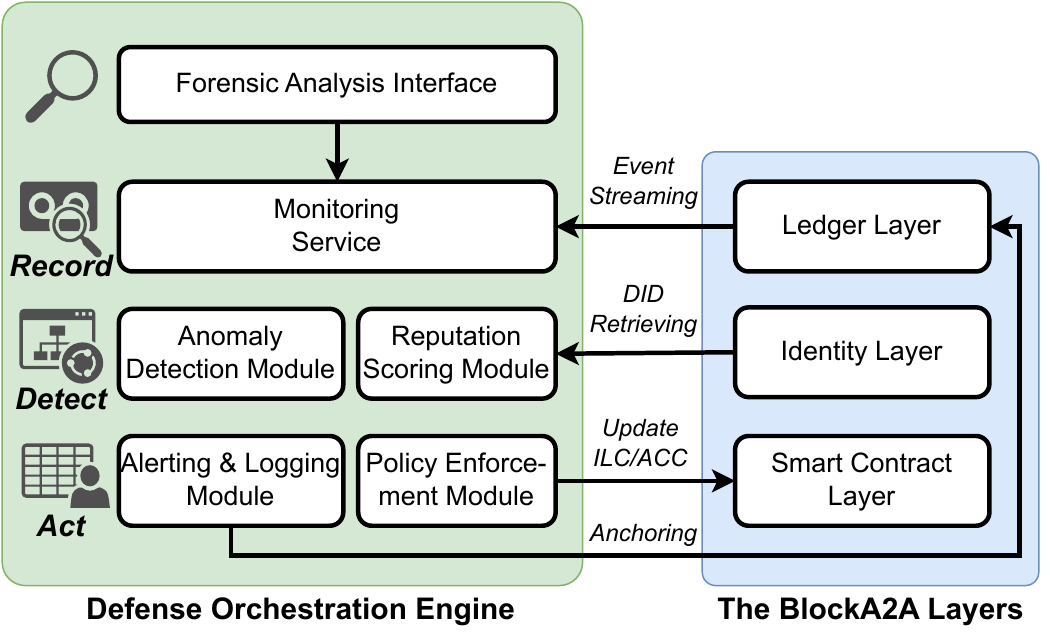}
    \caption{Defense Orchestration Engine Architecture}
    \label{fig:defense-architecture}
\end{figure}

\subsection{Functional Components}

The architectural design of the Defense Orchestration Engine is illustrated in Figure \ref{fig:defense-architecture}. This framework orchestrates multiple interconnected components to enable seamless threat management, from initial detection to post-incident analysis.

The \textit{Monitoring Service} continuously observes on-chain events, such as smart contract executions and task state transitions, as well as agent interactions recorded in the Ledger Layer. By connecting to the Provenance Smart Contract and Interaction Logic Contracts, it streams real-time data to generate event streams that are fed into the Anomaly Detection Module for analysis. This continuous data collection forms the foundation for identifying deviations from normal system behavior.

The \textit{Reputation Scoring Module} calculates and updates decentralized identifier (DID) reputation scores using a Bayesian model with decay factors to account for the recency of interactions. Scores are determined based on historical interaction success/failure rates, consensus participation and validation performance, and adherence to smart contract rules and protocols. These scores are stored as on-chain metadata associated with each DID, providing a dynamic measure of trustworthiness that informs security decisions across the ecosystem.

The \textit{Anomaly Detection Module} employs time-series analysis of message frequency and response times, state transition pattern recognition, and machine learning models for behavioral clustering to detect deviations from established baselines. These baselines are built using historical data from the Forensic Analysis Interface, which captures normal system behavior over time. When anomalies are identified, the module generates alerts with confidence scores, enabling timely intervention to address suspicious activities.

The \textit{Policy Enforcement Module} dynamically adjusts agent permissions and system behavior by interacting with Access Control Contracts (ACC) to modify authorization policies and Interaction Logic Contracts (ILC) to alter workflow execution rules. Triggered by alerts from the Anomaly Detection Module and reputation thresholds from the Reputation Scoring Module, this component executes smart contract transactions to enforce security policies, such as restricting access or modifying workflow logic, in response to emerging threats.

The \textit{Alerting and Logging Module} prioritizes alerts and maintains a comprehensive record of all security events and response actions. Alerts are disseminated through multiple channels, including on-chain event logs, off-chain notification services (e.g., email, SMS), and real-time dashboards, ensuring timely awareness for system administrators. Logs are stored in both on-chain immutable records and off-chain searchable databases, providing a reliable audit trail for post-incident analysis.

The \textit{Forensic Analysis Interface} provides tools for querying and analyzing on-chain audit trails to support post-incident investigations. Capabilities include time-series reconstruction of events, smart contract execution tracing, and DID interaction graph visualization. By leveraging the Data Anchoring Protocol, the interface verifies the integrity of historical data, ensuring the accuracy and reliability of forensic findings.

\begin{algorithm}
    \SetAlgoLined
    \KwIn{Reputation scores of agents, predefined threshold $\tau$}
    \KwOut{Off-chain evidence, on-chain log/alert, updated AGC}
    Identify agents with reputation score $< \tau$\;
    Collect evidence of suspicious behavior, \ie on-chain task records and off-chain task data, for flagged agents\;
    Update the agent's AGC to reflect the suspicious status\;
    Generate on-chain alert (agent DID, timestamp, reason) with off-chain evidence for the security team.
    \caption{Byzantine Agent Flagging Process}
    \label{alg:byzantine-flagging}
    \end{algorithm}
    
    \begin{algorithm}
    \SetAlgoLined
    \KwIn{Received data $D$, sender's DID $S$, on-chain hash $H_{\text{on-chain}}$}
    \KwOut{Off-chain evidence, on-chain log/alert, updated ILC guard}
    Compute hash $H_{\text{received}} = \text{Hash}(D)$\;
    \If{$H_{\text{received}} \neq H_{\text{on-chain}}$}{
        Create off-chain evidence of prompt tampering\;
        Update the ILC guard to halt task execution\;
        Generate tampering log/alert on chain.
    }
    \caption{Execution Halt Upon Prompt Tampering}
    \label{alg:tamper-detection}
    \end{algorithm}
    
    \begin{algorithm}
    \SetAlgoLined
    \KwIn{Agent permission list $P$, activity logs $L$, threat detection signal $\sigma$}
    \KwOut{Update ACC policies, on-chain log}
    \If{$\sigma = \text{True}$ AND suspicious activity detected in $L$}{
        Identify relevant resources/actions $P$ through AGC\;
        Update policies of the corresponding ACCs to revoke permissions\;
        Log revocation digests (agent ID, resource, action timestamp) on chain.
    }
    \caption{Real-time Permission Revocation}
    \label{alg:permission-revocation}
    \end{algorithm}

\subsection{Dynamic Counter-Attack Mechanisms}

The engine implements three principal countermeasure algorithms to neutralize active agent threats. The \textit{Byzantine Agent Flagging Process} (Algorithm \ref{alg:byzantine-flagging}) identifies agents with reputation scores below a predefined threshold, collects evidence of suspicious behavior, and generates alerts for the security team, which is logged on chain. This process enables rapid discovery of potentially malicious actors.

The \textit{Task Halt Upon Prompt Tampering} algorithm (Algorithm \ref{alg:tamper-detection}) verifies the integrity of received data by comparing its hash with the hash stored on chain. If a mismatch is detected, indicating potential tampering, the algorithm creates an off-chain evidence for the tampering, generates an on-chain alert, and updates the task's corresponding ILC's guards to halt task execution, preventing the propagation of corrupted data.

The \textit{Real-time Permission Revocation} mechanism (Algorithm \ref{alg:permission-revocation}) checks agent permissions and suspicious activity patterns to revoke access to resources/actions bound to a specific agent when a threat is detected. Upon detection of such a threat, the algorithm first identifies relevant resources/actions (\ie the permission list $P$) through the suspicious agent's AGC. Further, by updating the policies of the corresponding ACCs and logging revocation digests on chain, the algorithm ensures immediate isolation of malicious agents to mitigate ongoing attacks.

\subsection{Interaction with BlockA2A Layers}

The DOE integrates deeply with each layer of the BlockA2A architecture to ensure cohesive security operations. At the \textit{Identity Layer}, it uses DID documents and reputation scores to authenticate and authorize agents, while Cross-Chain DID Validation facilitates interoperable threat intelligence sharing across different MASes or blockchain networks (\eg the same agent involved in multiple MASes). At the \textit{Ledger Layer}, the engine monitors Provenance Smart Contracts for task state transitions and uses the Data Anchoring Protocol to maintain tamper-proof logs of security events, ensuring the integrity and immutability of audit data. Within the \textit{Smart Contract Layer}, the DOE dynamically updates Access Control Contracts (ACC) and Interaction Logic Contracts (ILC) to enforce permissions and adjust workflow execution during threats. This integration enables real-time policy enforcement and adaptive response to evolving security challenges.




\section{Evaluation}
\label{sec:evaluation}

In this section, we perform extensive empirical studies and experimental evaluations to assess \sys's effectiveness in thwarting agent-to-agent collaboration threats and its operational efficiency. Specifically, we first carry out an in-depth case study in analyzing how \sys (combined with the DOE) uniquely enables the defense of various types of MAS attacks. After that, we evaluate the additional overhead required to instantiate \sys within an existing MAS, as well as the operational costs of the DOE. Besides, we also provide a detailed routine for integrating \sys into Google's A2A~\cite{a2a-protocol} following our instantiation framework (\S~\ref{sec:instantiation-framework}), demonstrating its practicality. Experimental results show that \sys introduces reasonable additional overhead to existing A2A frameworks while offering excellent performance in locating malicious agents/prompts and enforcing defense policies.

\subsection{Empirical Study: Defense Effectiveness Analysis}

This subsection presents an empirical analysis of the Defense Orchestration Engine's capability to mitigate four categories of agent-to-agent system threats (mentioned in \S~\ref{sec:literature-review}) supported by a structured evaluation of attack vectors and corresponding defense mechanisms. The analysis leverages the DOE's integration with \sys's three-layer architecture to demonstrate resilience against emerging threats in multi-agent systems (MAS).

\subsubsection{Threat Categories and Defense Mechanisms}

Table \ref{tab:attack-defense-compact} summarizes the evaluated threats, their technical characteristics, and how we employ DOE's components to neutralize them. Each defense strategy is rooted in the DOE's modular design, combining on-chain monitoring, cryptographic integrity checks, dynamic policy enforcement, and reputation-based trust management.

\parab{Prompt-Based Attacks}
Prompt-based attacks exploit language model vulnerabilities through adversarial input manipulation. For example, \textit{jailbreak attacks} \cite{jailbreak-survey} use crafted prompts to bypass ethical safeguards, while \textit{prompt injection} \cite{g-safeguard} introduces malicious instructions to induce misinformation. The most advanced variant, \textit{prompt infection} \cite{netsafe, prompt-infection}, propagates self-replicating malicious prompts across agents, akin to a cyber virus. 

\textit{DOE Defense:}  
The DOE mitigates these threats through a multi-layered approach:
\begin{itemize}
    \item Cryptographic Integrity Checks: The Data Anchoring Protocol (Protocol~\ref{prot:data-anchoring}) stores SHA-256 hashes of valid prompts on the Ledger Layer, enabling real-time verification of received content against on-chain records. Merkle trees are used for large-scale prompt datasets, allowing efficient detection of tampered or replicated content.
    \item Dynamic Isolation: Upon detecting hash mismatches (indicating tampering), the Policy Enforcement Module revokes the sender's communication privileges via the Access Control Contract (ACC) and quarantines the agent based on reputation scores.
    \item Immutable Traceability: Crucially, the DOE leverages the Ledger Layer's Provenance Smart Contract to log all prompt interactions with timestamps and DID signatures. This creates an immutable audit trail that enables backward tracing of malicious prompts to their origin. For example:
\begin{itemize}

    \item Jailbreak attacks can be traced to specific DIDs by cross-referencing unauthorized prompt hashes with transaction logs.
    \item Prompt injection incidents are attributed to senders via BLS signatures anchored on-chain, while timestamp ordering reveals propagation sequences.
    \item Prompt infections are mapped using Merkle tree-based propagation graphs, allowing the Forensic Analysis Interface to identify the first agent to introduce the malicious prompt and model its spread across the ecosystem.
This traceability transforms the blockchain into a forensic tool, enabling post-incident attribution, validation of attack chains, and proactive blocking of repeat offenders through reputation-based sanctions.
\end{itemize}
\end{itemize}

\parab{Communication-Based Attacks}
Communication-based attacks target inter-agent message integrity and availability. Examples include:  
\textit{Agent-in-the-Middle (AiTM) attacks} \cite{red-team-comm}, which intercept and alter messages;  
\textit{False-data injection} \cite{ripple-effect}, disrupting control processes by injecting fraudulent data;  
\textit{Contagious recursive blocking} \cite{corba}, overwhelming agents with repetitive requests.  

\textit{DOE Defense:}  
The Monitoring Service continuously tracks message flow patterns in the Ledger Layer's Provenance Smart Contract. For AiTM attacks, the system uses message signatures (from the Identity Layer) to verify sender authenticity and detects unauthorized interceptions via timestamp anomalies. False-data injection is mitigated by cross-referencing data with oracles (via the Authenticated Data Import Protocol) and flagging inconsistencies. Recursive blocking attacks are identified by the Anomaly Detection Module through message frequency thresholds; the Policy Enforcement Module then applies rate limiting via the Interaction Logic Contract (ILC), temporarily suspending overactive agents.

\parab{Behavioral/Psychological Attacks}
These attacks exploit agent decision-making processes, such as: \textit{Dark personality manipulation} \cite{psysafe}, where agents with simulated malicious traits induce risky actions;  \textit{Malfunction amplification} \cite{breaking-agents}, misleading agents into redundant tasks.  

\textit{DOE Defense:}  
The Reputation Scoring Module maintains behavioral profiles using Bayesian modeling, flagging agents with deviations from expected ethical norms (e.g., high failure rates in consensus participation). The Forensic Analysis Interface reconstructs decision chains to identify manipulation patterns, while the Policy Enforcement Module enforces contextual access controls via ACCs, restricting agents from high-risk actions when suspicious behavior is detected. For example, agents with "dark personality" traits are denied access to sensitive resources until their reputation scores improve.

\parab{Systemic/Architectural Attacks}
These attacks target MAS infrastructure, including: \textit{Topological vulnerability exploitation} \cite{g-safeguard, achilles-heel}, leveraging network structures for misinformation propagation; \textit{Distributed safety bypass} \cite{agents-under-siege}, exploiting latency to evade detection.  

\textit{DOE Defense:}  
The Anomaly Detection Module uses machine learning to identify topological attack patterns (e.g., sudden spikes in misinformation propagation). The Policy Enforcement Module dynamically reconfigures ILC workflows to introduce delay-tolerant validation checkpoints, countering latency-based exploits. Cross-chain threat intelligence sharing via the Cross-Chain DID Validation Protocol enhances collective defense against systemic attacks, while the Data Anchoring Protocol ensures immutable logging of architectural changes for post-incident analysis.

\begin{table*}[ht]
    \centering
    \begin{adjustbox}{max width=\textwidth}
    \begin{tabular}{|c|c|c|c|}
        \hline
        \textbf{Threat Category} & \textbf{Specific Attack} & \textbf{Technical Mechanism} & \textbf{DOE Defense Strategy} \\
        \hline
        \multirow{3}{*}{Prompt-Based} 
            & Jailbreak Attacks & \makecell{Adversarial prompts\\bypass LLM safeguards \cite{jailbreak-survey}} & \makecell{On-chain hash anchoring;\\ACC blocks unauthorized patterns} \\
            \cline{2-4}
            & Prompt Injection & \makecell{Malicious instructions\\induce harmful outputs \cite{g-safeguard}} & \makecell{Real-time hash verification;\\agent isolation} \\
            \cline{2-4}
            & Prompt Infection & \makecell{Self-replicating\\malicious prompts \cite{netsafe}} & \makecell{Merkle tree checks;\\reputation quarantine} \\
        \hline
        \multirow{3}{*}{Communication-Based} 
            & Agent-in-the-Middle (AiTM) & \makecell{Message interception/\\tampering \cite{red-team-comm}} & \makecell{Message signature verification;\\timestamp anomalies} \\
            \cline{2-4}
            & False-Data Injection & \makecell{Fraudulent data\\disrupts processes \cite{ripple-effect}} & \makecell{Oracle-backed validation;\\ILC state guards} \\
            \cline{2-4}
            & Contagious Recursive Blocking & \makecell{Repetitive messages\\overwhelm agents \cite{corba}} & \makecell{Message throttling via ILC;\\reputation penalties} \\
        \hline
        \multirow{2}{*}{Behavioral/Psychological} 
            & Dark Personality Manipulation & \makecell{Exploits simulated\\malicious traits \cite{psysafe}} & \makecell{Bayesian reputation modeling;\\ACC restrictions} \\
            \cline{2-4}
            & Malfunction Amplification & \makecell{Misleads into\\redundant tasks \cite{breaking-agents}} & \makecell{Workflow trace analysis;\\dynamic task prioritization} \\
        \hline
        \multirow{2}{*}{Systemic/Architectural} 
            & Topological Vulnerability Exploitation & \makecell{Misinformation via\\network structures \cite{g-safeguard}} & \makecell{ML-based anomaly detection;\\cross-chain sharing} \\
            \cline{2-4}
            & Distributed Safety Bypass & \makecell{Latency exploits\\evasion detection \cite{agents-under-siege}} & \makecell{Delay-tolerant checkpoints;\\on-chain audits} \\
        \hline
    \end{tabular}
    \end{adjustbox}
    \caption{Compact Attack Vector Analysis and DOE Defense Mechanisms}
    \label{tab:attack-defense-compact}
\end{table*}



In summary, our empirical study results highlight the DOE's capability to address diverse threats by integrating cryptographic safeguards, behavioral analytics, and dynamic policy enforcement across \sys's architectural layers.

\subsection{Instantiation of \sys within Google A2A}
\label{subsec:google-a2a-instantiation}

This section presents a in-detail walk-through to instantiate \sys within Google A2A, enhancing its accountability, traceability, and security, which adheres to our \sys instantiation framework. This instantiation integrates \sys's Identity, Ledger, and Smart Contract layers with Google A2A's core components.

\subsubsection{Transformation Functions for Google A2A}

For Google A2A, we define the following transformation functions to bridge its protocol to \sys:

\begin{enumerate}

\item \textit{Identity Mapping Function (\(\mathcal{I}_{\text{G-A2A}}\))}: Map Google A2A agent identifiers to DIDs. For example, an agent's service point, Google Cloud project ID and service account are combined to form a DID: \textsf{did:web:agent.example.com:gcp-project-123}. Meanwhile, $\mathcal{I}_{\text{G-A2A}}$ constructs a DID document based on A2A's AgentCard and binds it to the DID (\eg populating the \textsf{capabilities} field of agent DID document based on the \textsf{skills} field of AgentCard).

\item \textit{Metadata Transformation (\(\mathcal{T}_{\text{G-A2A}}\))}: Transforms Google A2A task metadata (including task ID, context ID, task status) into \sys's DID-driven task specifications, including the client's DID, the server's DID and blockchain-achored payload hashes.

\item \textit{Protocol Translation (\(\mathcal{A}_{\text{G-A2A}}\))}: Google A2A's JSON-RPC-based client-server communication are enhanced by sender/recipient DIDs and anchored as blockchain transactions. For example, a message/send request is converted into a signed transaction recording the message sender (DID), recipient (DID), and content hash on the ledger. At the same time, task status update (e.g., in Streaming Task Execution and Multi-Turn Interaction) is now recorded as a state-transition blockchain transaction, following Protocol~\ref{prot:state-validation}. 

\end{enumerate}

\subsubsection{Layer-Specific Instantiation}

\parab{Identity Layer Instantiation} 
This instantiation enhances Google A2A's centralized TLS-based authentication with DIDs, enabling cross-platform identity verification:

\begin{itemize}

\item \(\mathcal{I}_{\text{G-A2A}}\): Map Google A2A agent IDs to DIDs as described above;

\item \(\mathcal{S}_{\text{G-A2A}}\): Sign Google A2A messages using the agent's DID private key, appending a DID signature to the message header;

\item \(\mathcal{V}_{\text{G-A2A}}\): Verify messages by resolving the sender's DID to fetch the public key and validate the signature;

\item \(\mathcal{R}_{\text{G-A2A}}\): Resolve DIDs to DID documents, which are enriched by Google A2A AgentCards.

\end{itemize}

When sending a message via message/send, the Google A2A agent uses \(\mathcal{S}_{\text{G-A2A}}\) to sign the message with its DID key, and the recipient uses \(\mathcal{V}_{\text{G-A2A}}\) to verify the signature via DID resolution. This serves as an effective enhancement or replacement of the Google A2A's HTTPS-driven message authentication.

\parab{Ledger Layer Instantiation}
The Ledger Layer anchors all Google A2A interactions, providing a tamper-proof audit trail for compliance and forensics:

\begin{itemize}

\item \(\mathcal{T}_{\text{task-GA}}\): Transforms Google A2A task metadata (e.g., from tasks/get responses) into \sys task records, including initiator DID, participant DIDs, and timestamps.

\item \(\mathcal{A}_{\text{state-GA}}\): Converts Google A2A task state transitions (e.g., submitted to completed) into on-chain transactions, recording each step with a block timestamp.

\item \(\mathcal{A}_{\text{anchor-GA}}\): Anchors large Google A2A artifacts (e.g., files sent via FilePart) by storing their hashes on the ledger and referencing IPFS CIDs for off-chain storage.

\end{itemize}

When a Google A2A task transitions to completed, \(\mathcal{A}_{\text{state-GA}}\) generates a transaction recording the state change, the final artifact hashes, and the agent DIDs involved. Large output files are stored on IPFS, with their CIDs anchored on the ledger via \(\mathcal{A}_{\text{anchor-GA}}\).

\parab{Smart Contract Layer Instantiation}
\sys's Smart Contract Layer works as an out-of-box add-on to Google A2A, compensating for its lack of resource authorization, complex task lifecycle management and agent governance.

\begin{itemize}

\item \(\mathcal{C}_{\text{ACC-GA}}\): Generates Access Control Contracts (ACC) according to server-side or client-side resource authorization requirements, mapping them to DID-based access rules.

\item \(\mathcal{C}_{\text{ILC-GA}}\): Creates Interaction Logic Contracts (ILC) for complex Google A2A task workflows (e.g., involving multi-round communication among multiple agents), automating state transitions and multi-party approvals.

\item \(\mathcal{C}_{\text{ARC-GA}}\): Develops Agent Capability Registries (ARC) by translating Google A2A's skills in agent cards into on-chain capability definitions.

\end{itemize}

A Google A2A agent's skills are converted via \(\mathcal{C}_{\text{ARC-GA}}\) into an ARC that defines the capability's input/output formats, access policies, and execution rules on the blockchain. Access to this skill is controlled by an ACC that verifies the requester's DID against allowed attributes (e.g., organization membership) and context information. For complex tasks requiring multi-agent participation and multi-stage execution, ILCs enforce automated agent interaction flows with native task traceability.

By applying the \sys instantiation framework, Google A2A interactions inherit the following trust properties:

\begin{enumerate}

\item \textbf{Authenticity}: Google A2A agents are verified via DIDs, ensuring that messages originate from claimed identities.

\item \textbf{Integrity}: All task metadata and state transitions are cryptographically anchored on the ledger, preventing tampering.

\item \textbf{Accountability}: Every action (e.g., message send, task update) is linked to a DID, enabling end-to-end traceability of agent interactions.

\end{enumerate}

The trust preservation is guaranteed by the Cross-MAS Trust Equivalence Theorem (Theorem~\ref{theorem:trust-equivalence}), as the transformation functions rigorously map Google A2A protocols to \sys's verifiable transactions. This instantiation enables Google A2A to leverage \sys's decentralized trust infrastructure while maintaining compatibility with its existing communication protocols, thus enhancing accountability, traceability, and security without disrupting operational workflows.

\begin{table*}[h]
\centering
\caption{Operational Costs of \sys Components (Average Latency)}
\label{tab:operational-costs}
\begin{adjustbox}{max width=\columnwidth}
\begin{tabular}{l l r r}
\toprule
\textbf{Category} & \textbf{Operation} & \textbf{Duration (ms)} & \textbf{Operation Type} \\
\midrule
\multirow{2}{*}{DID Registration} 
    & On-chain hash anchoring & 27.9 & On chain \\
    & Off-chain document storage & 7.5 & Off chain \\
\addlinespace

\multirow{3}{*}{Message Authentication}
    & Signature generation & 0.2 & Off chain \\
    & DID document retrieval & 2.3 & Off chain \\
    & Signature verification & 13.0 & Off chain \\
\addlinespace

\multirow{3}{*}{Task Recording}
    & Task initialization & 35.0 & On chain \\
    & Multi-signature \& Public key aggregation & 45.4 & On\&Off chain \\
    & State transition & 64.0 & On chain \\
    & Data anchoring & 34.9 & On chain \\
\addlinespace

\multirow{2}{*}{Access Control}
    & ACC token issuance & 26.5 & On chain \\
    & Off-chain token verification & 7.0 & Off chain \\
\bottomrule
\end{tabular}
\end{adjustbox}
\end{table*}

\subsection{Operational Cost of \sys}
\label{subsec:cost-evaluation}


To test whether \sys integration works in practice, this section measures the extra overhead it adds to existing MASes. We focus on computational costs (including memory-access latency), avoiding network factors like bandwidth or latency, as these depend heavily on how the blockchain and IPFS are deployed. Therefore, we run all experiments locally on a Linux server with a multi-core x86\_64 Intel CPU (2.60 GHz), using default setups for a Hardhat testnet and IPFS Kubo. Cryptographic operations (SHA-256, ECDSA secp256k1, and BLS BLS12-381) use standard implementations. Each measurement is averaged over 10 runs. Table~\ref{tab:operational-costs} shows the latency for key \sys operations, grouped into four categories: DID registration, message authentication, task recording, and access control.

\noindent Key observations from our evaluation are summarized below:

\begin{itemize}
    \item \textbf{Low-impact off-chain operations}: Off-chain operations, including IPFS interaction (document storage and retrieval) and token/signature verification introduce negligible overhead (mostly $<$10ms), in par with conventional cryptographic operations in MASes.
    
    \item \textbf{Moderate-cost on-chain operations}: Operations with blockchain interaction need tens of milliseconds for computation, which are more significant than that of off-chain operations but still in a moderate level. Moreover, these operations typically occur infrequently (\eg during agent registration, task initialization or milestone completion).
    
    \item \textbf{Cost mitigation}: The highest latencies occur only during critical trust points, \ie state transitions which require pairing-based multi-signature verification on chain. Nevertheless, further aggregation and batched verification of multiple signatures (for multiple transitions) are feasible for latency-sensitive tasks.
    
    \item \textbf{Storage efficiency}: Off-chain storage (IPFS/cloud) handles over 92\% of payload data, limiting on-chain costs to metadata hashes.
\end{itemize}

In summary, the operational costs introduced by \sys remain within practical boundaries for most MAS deployments, considering seconds-level LLM inferences in a MAS. The maximum observed overhead (64.0ms for state transitions) is in still lower than typical MAS task durations, while providing verifiable non-repudiation and tamper-evidence. In real-world deployment, the transactional latencies of on-chain operations and network conditions for interacting with IPFS/cloud storage could be additional bottlenecks of \sys performance. Nevertheless, the framework's modular design allows MAS architects to selectively deploy high-value components where security requirements justify the marginal resource investment.

\begin{table*}[h]
\centering
\caption{Response Time of DOE Counter-Attack Algorithms (Average Latency)}
\label{tab:doe-response-times}
\begin{tabular}{l l r r}
\toprule
\textbf{Algorithm} & \textbf{Operation} & \textbf{Duration (ms)} & \textbf{Operation Type} \\
\midrule
\multirow{3}{*}{\shortstack{Byzantine Agent\\Flagging}}
    & Evidence collection & 53.7 & Off chain \\
    & AGC status update & 55.8 & On chain \\
    & On-chain alert generation & 25.0 & On chain \\
    \textbf{Total} & & \textbf{135.0} & \\
\addlinespace

\multirow{3}{*}{\shortstack{Execution Halt\\Upon Tampering}}
    & Tamper evidence storage & 8.0 & Off chain \\
    & ILC guard update & 53.3 & On chain \\
    & On-chain alert generation & 25.0 & On chain \\
    \textbf{Total} & & \textbf{87.0} & \\
\addlinespace

\multirow{3}{*}{\shortstack{Real-time\\Permission Revocation}}
    & AGC capability resolution & 6.8 & On chain \\
    & ACC policy update & 57.5 & On chain \\
    & On-chain revocation log & 27.5 & On chain \\
    \textbf{Total} & & \textbf{91.8} & \\
\bottomrule
\end{tabular}
\end{table*}

\subsection{Response Timeliness of DOE}
\label{subsec:doe-response-eval}

To assess the reactivity and timeliness of the Defense Orchestration Engine's (DOE) counter-attack mechanisms, we measured the average execution latency of its three principal algorithms from repetitive experiments over our prototype implementation. Table~\ref{tab:doe-response-times} demonstrates that all critical security operations complete within sub-second timeframes, enabling effective neutralization of active threats.

\noindent In particular, we find from the experimental results that:

\begin{itemize}
    \item \textbf{Sub-second neutralization}: All three counter-attack algorithms complete within $\sim$110ms on average, with critical-path operations (ILC guard updates, ACC policy changes) executing in 60ms.
    
    \item \textbf{Parallel off-chain evidence collection with on-chain action}: While our current implementation execute the operations in the same algorithms consecutively, in real-world deplotment, evidence collection can operate concurrently with on-chain security actions, adding minimal latency to threat containment.

    \item \textbf{Scalability Potential}: While Layer-2 blockchain solutions can futher reduce blockchain transaction time, concurrent processing of evidence collection across multiple nodes can improve throughput for high-volume environments.
\end{itemize}

In conclusion, the DOE's counter-attack algorithms demonstrate highly reactive performance with sub-second containment capabilities. The maximum observed latency (135.0ms) represents less than 10\% of the seconds-level LLM-based agent tasks, proving the framework's capability to neutralize attacks before significant damage occurs. Moreover, this efficiency exhibits further-optimization potential through optimized blockchain interactions, parallelized evidence collection and on-chain critical actions - making DOE suitable for real-time defense in sensitive MAS environments.
\section{Conclusion}

Agent-to-agent collaboration is pivotal to the transformative potential of agentic AI, yet existing frameworks remain vulnerable to evolving threats due to centralized trust models, brittle auditability, and static security policies. \sys addresses these gaps by unifying decentralized identity, blockchain-anchored audit trails, and smart contract-driven policies to establish a scalable, trustless foundation for secure collaboration. Our evaluation confirms its strong defensive capabilities against a wide range of threats and demonstrates its seamless integration with existing frameworks like Google A2A. Moreover, \sys operates with minimal overhead and sub-second response times, making it a viable and impactful defense mechanism for real-time MAS security. By reconciling security with flexibility, \sys not only mitigates current risks but also provides a modular substrate for future defenses, paving the way for resilient, enterprise-scale AI ecosystems where autonomous agents collaborate safely across organizational and technical boundaries.


\bibliographystyle{plain}
\bibliography{./bibliography/refs}


\end{document}